\documentclass[a4paper,UKenglish,cleveref,autoref,thm-restate]{lipics-v2019}

\usepackage[utf8]{inputenc}
\usepackage[T1]{fontenc}

\usepackage{adjustbox}
\usepackage{booktabs}
\usepackage{listings}
\usepackage{multicol}
\usepackage[numbers]{natbib}
\usepackage[autolanguage]{numprint}
\usepackage{proof}
\usepackage{softdev}
\usepackage{subcaption}
\usepackage{wrapfig}
\usepackage{xspace}

\usepackage[ruled]{algorithm2e} 

\SetAlFnt{\small}
\SetAlCapFnt{\small}
\SetAlCapNameFnt{\small}
\SetAlCapHSkip{0pt}
\IncMargin{-\parindent}

\let\origthelstnumber\thelstnumber
\makeatletter
\newcommand*\Suppressnumber{%
  \lst@AddToHook{OnNewLine}{%
    \let\thelstnumber\relax%
     \advance\c@lstnumber-\@ne\relax%
    }%
}

\newcommand*\Reactivatenumber[1]{%
  \setcounter{lstnumber}{\numexpr#1-1\relax}
  \lst@AddToHook{OnNewLine}{%
   \let\thelstnumber\origthelstnumber%
   \refstepcounter{lstnumber}
  }%
}

\newcommand{\cpctplus}{\textrm{\textit{CPCT}}$^+$\xspace}
\newcommand{\cpctplusdontmerge}{\textrm{\textit{CPCT$^+_\textrm{DM}$}}\xspace}

\newcommand{\cpctplusrev}{\textrm{\textit{CPCT$^+_\textrm{rev}$}}\xspace}
\newcommand{\crarrow}{\rightarrow_{\textrm{\tiny CR}}}
\newcommand{\crarrowstar}{\rightarrow_{\textrm{\tiny CR}}^*}

\newcommand{\lrarrowstar}{\rightarrow_{\textrm{\tiny LR}}^*}
\newcommand{\lrarrow}{\rightarrow_{\textrm{\tiny LR}}}
\newcommand{\mf}{\textrm{\textit{MF}}\xspace}

\newcommand{\panic}{\textrm{Panic mode}\xspace}

\newcommand{\aho}{Aho and Peterson\xspace}
\newcommand{\corchuelo}{Corchuelo \emph{et al.}\xspace}
\newcommand{\fischer}{Fischer \emph{et al.}\xspace}
\newcommand{\holub}{Holub\xspace}

\newcommand{\cpctplusreverrorlocsratioovercpctplus}{31.93\%{\footnotesize$\pm$0.289\%}\xspace}

\newcommand{\numruns}{\numprint{30}\xspace}
\newcommand{\numbootstrap}{\numprint{10000}\xspace}
\newcommand{\corpussize}{\numprint{200000}\xspace}
\newcommand{\corpussizemb}{\numprint{401}\xspace}
\newcommand{\cpctplussuccessrate}{98.37\%{\footnotesize$\pm$0.017\%}\xspace}
\newcommand{\cpctplusfailurerate}{1.63\%{\footnotesize$\pm$0.017\%}\xspace}

\newcommand{\cpctpluserrorlocs}{\numprint{435812}{\footnotesize$\pm$\numprint{473}}\xspace}

\newcommand{\cpctpluslongerfailurerate}{1.02\%{\footnotesize$\pm$0.016\%}\xspace}

\newcommand{\panicerrorlocs}{\numprint{981628}{\footnotesize$\pm$\numprint{0}}\xspace}

\title{Don't Panic! Better, Fewer, Syntax Errors for LR Parsers}

\author{Lukas Diekmann}{Software Development Team, King's College London, United Kingdom \and \url{https://lukasdiekmann.com/}}{lukas.diekmann@gmail.com}{}{}
\author{Laurence Tratt}{Software Development Team, King's College London, United Kingdom \and \url{https://tratt.net/laurie/}}{laurie@tratt.net}{https://orcid.org/0000-0002-5258-3805}{}
\authorrunning{L.\,Diekmann and L.\,Tratt}
\Copyright{Lukas Diekmann and Laurence Tratt}
\relatedversion{Updates will be made available at \url{https://arxiv.org/abs/1804.07133}}

\EventEditors{Robert Hirschfeld and Tobias Pape}
\EventNoEds{2}
\EventLongTitle{34th European Conference on Object-Oriented Programming (ECOOP 2020)}
\EventShortTitle{ECOOP 2020}
\EventAcronym{ECOOP}
\EventYear{2020}
\EventDate{July 13--17, 2020}
\EventLocation{Berlin, Germany}
\EventLogo{}
\SeriesVolume{166}
\ArticleNo{6}

\nolinenumbers

\begin{document}

\maketitle

\begin{abstract}
Syntax errors are generally easy to fix for humans, but not for parsers in general
nor LR parsers in particular. Traditional `panic mode' error recovery, though
easy to implement and applicable to any grammar,
often leads to a cascading chain of
errors that drown out the original. More advanced error recovery techniques
suffer less from this problem but have seen little practical use because
their typical performance was seen as poor, their worst case unbounded, and the
repairs they reported arbitrary. In this paper we introduce the \cpctplus algorithm,
and an implementation of that algorithm, that address these issues. First, \cpctplus
reports the complete set of minimum cost repair sequences for a given location, allowing
programmers to select the one that best fits their intention. Second, on a
corpus of \corpussize real-world syntactically invalid Java programs, \cpctplus is
able to repair \cpctplussuccessrate of files within
a timeout of 0.5s. Finally, \cpctplus uses the complete set of
minimum cost repair sequences to reduce the cascading error problem, where
incorrect error recovery causes further spurious syntax errors to be identified.
Across the test corpus, \cpctplus reports
\cpctpluserrorlocs error locations to the user, reducing the cascading
error problem substantially relative to the \panicerrorlocs error locations reported by panic mode.
\end{abstract}

\section{Introduction}

Programming is a humbling job which requires acknowledging that we will make
untold errors in our quest to perfect a program. Most troubling are semantic
errors, where we intended the program to do one thing, but it does another. Less
troubling, but often no less irritating, are syntax errors, which are
generally minor deviances from the exacting syntax required by a compiler.
So common are syntax errors that parsers in modern compilers are designed
to cope with us making several: rather than stop on the first syntax error, they attempt
to \emph{recover} from it. This allows them to report, and us to fix, all our
syntax errors in one go.

When error recovery works well, it is a useful productivity gain. Unfortunately,
most current error recovery approaches are simplistic. The most common
grammar-neutral approach to error recovery are those algorithms described as
`panic mode' (e.g.~\cite[p.~348]{holub90compilerdesign}) which skip
input until the parser finds something it is able to parse. A more
grammar-specific variation of this idea is to skip input until a pre-determined
synchronisation token (e.g. `;' in Java) is
reached~\cite[p.~3]{degano95comparison}, or to try inserting a single
synchronisation token. Such strategies are often unsuccessful,
leading to a
cascade of spurious syntax errors (see
Figure~\ref{fig:javaerror} for an example). Programmers quickly learn that
only the location of the first error in a file -- not the reported repair, nor the location of
subsequent errors -- can be relied upon to be accurate.

\begin{figure}[t]
\begin{minipage}{0.49\textwidth}
\begin{tabular}{p{0.02\textwidth}p{0.45\textwidth}}
\begin{subfigure}{0.02\textwidth}
\caption{}
\label{lst:javaerror:input}
\end{subfigure}
&
\begin{minipage}[t]{0.86\textwidth}
\vspace{-13.5pt}
\begin{lstlisting}[language=Java]
class C {
  int x y;
}
\end{lstlisting}
\end{minipage}
\\
\begin{subfigure}{0.02\textwidth}
\addtocounter{subfigure}{1}
\caption{}
\label{lst:javaerror:cpctplus}
\end{subfigure}
&
\begin{minipage}[t]{0.86\textwidth}
\vspace{-13.5pt}
\begin{lstlisting}
Parsing error at line 2 col 9. Repair
sequences found:
  1: Delete y
  2: Insert ,
  3: Insert =
\end{lstlisting}
\end{minipage}
\end{tabular}
\end{minipage}
\begin{minipage}{0.49\textwidth}
\vspace{-27pt}
\begin{tabular}{p{0.02\textwidth}p{0.48\textwidth}}
\begin{subfigure}{0.02\textwidth}
\addtocounter{subfigure}{-2}
\caption{}
\label{lst:javaerror:javac}
\end{subfigure}
&
\begin{minipage}[t]{0.9\textwidth}
\vspace{-13.5pt}
\begin{lstlisting}
C.java:2: error: ';' expected
  int x y;
       ^
C.java:2: error: <identifier> expected
  int x y;
         ^
\end{lstlisting}
\end{minipage}
\end{tabular}
\end{minipage}
\vspace{-10pt}
\caption{An example of a simple, common Java syntax error
(\subref{lst:javaerror:input}) and the problems traditional error recovery has in
dealing with it. \texttt{javac} (\subref{lst:javaerror:javac}) spots the error
when it encounters `\texttt{y}'. Its error recovery algorithm then
repairs the input by inserting a semicolon before `\texttt{y}' (i.e.~making
the input equivalent to `\texttt{int x; y;}'). This then causes a spurious
parsing error, since `\texttt{y}' on its own is not a valid statement. The \cpctplus
error recovery algorithm we introduce in this paper produces
the output shown in (\subref{lst:javaerror:cpctplus}): after spotting an error
when parsing encounters `\texttt{y}', it uses the Java grammar to find the
complete set of minimum cost repair sequences (unlike previous approaches which
non-deterministically find one minimum cost repair sequence). In this case three
repair sequences are reported to the user: one can delete `\texttt{y}'
entirely (`\texttt{int x;}'), or insert a comma
(`\texttt{int x, y;}'), or insert an equals sign (\texttt{`int x = y;'}).}
\label{fig:javaerror}
\end{figure}

It is possible to hand-craft error recovery algorithms for a specific language.
These generally allow better recovery from errors, but are challenging
to create. For example, the Java error recovery approach in the Eclipse IDE is 5KLoC long,
making it only slightly smaller than a modern version of Berkeley Yacc --- a
complete parsing system! Unsurprisingly, few real-world parsers contain
effective hand-written error recovery algorithms.

Most of us are so used to these trade-offs (cheap generic algorithms and poor
recovery vs.~expensive hand-written algorithms and reasonable recovery) that we
assume them to be inevitable. However, there is a long line of work on
more advanced generic error recovery algorithms. Probably the earliest such algorithm
is \aho~\cite{aho72minimum}, which, upon encountering an error, creates on-the-fly an
alternative (possibly ambiguous) grammar which allows the parser to recover.
This algorithm has fallen out of favour in programming language
circles, probably because of its implementation complexity and the difficulty of
explaining to users what recovery has been used. A simpler family of algorithms, which
trace their roots to \fischer~\cite{fischer79locally}, instead try to find a single minimum cost
\emph{repair sequence} of token insertions and deletions which allow the parser to
recover. Algorithms in this family are much better at recovering from errors
than naive approaches and can communicate the repairs they find in a way that
humans can easily replicate. However, such algorithms
have seen little practical use because their typical
performance is seen as poor and their worst case unbounded \cite[p.~14]{mckenzie95error}.
We add a further complaint: such approaches only report a single
repair sequence to users. In general -- and especially in syntactically rich
languages -- there are multiple reasonable repair sequences for a given error
location, and the algorithm has no way of knowing which best matches the user's intentions.

In this paper we introduce a new error recovery algorithm in the \fischer
family, \cpctplus. This takes the approach of \corchuelo~\cite{corchuelo02repairing}
as a base, corrects it, expands it, and optimises its implementation.
\cpctplus is simple to implement (under 500 lines of Rust code), is able to repair
nearly all errors in reasonable time, reports the
complete set of minimum cost repair sequences to users, and does so
in less time than \corchuelo.

We validate \cpctplus on a corpus of \corpussize real,
syntactically incorrect, Java programs (Section~\ref{experiment}). \cpctplus is
able to recover \cpctplussuccessrate of files within a 0.5s timeout and does
so while reporting fewer than
half the error locations as a traditional panic mode algorithm:
in other words, \cpctplus substantially reduces the cascading error problem.
We also show -- for, as far as we know, the first time -- that advanced error
recovery can be naturally added to a Yacc-esque system, allowing users
to make fine-grained decisions about what to do when error
recovery has been applied to an input (Section~\ref{sec:api}). We believe that this
shows that algorithms such as \cpctplus are ready for wider
usage, either on their own, or as part of a multi-phase recovery system.

\subsection{Defining the problem}

Formally speaking, we first test the following hypothesis:

\begin{description}
  \item[H1] The complete set of minimum cost repair sequences can
    be found in acceptable time.
\end{description}

The only work we are aware of with a similar concept of `acceptable time'
is~\cite{deJonge12natural}, who define it as the total time spent in error
recovery per file, with a threshold of 1s. We use that definition with one
change: Since many compilers are able to
fully execute in less than 1s, we felt that a tighter threshold is more
appropriate: we use 0.5s since we think that even the most
demanding users will tolerate such a delay. We strongly validate this hypothesis.

The complete set of minimum cost repair sequences makes it more likely that
the programmer will see a repair sequence that matches their original
intention (see Figure~\ref{fig:javaerror} for an example; Appendix~\ref{app:examples}
contains further examples in Java, Lua, and PHP). It also opens up a
new opportunity. Previous error recovery
algorithms find a single repair sequence, apply that to the input, and then
continue parsing. While that repair sequence may have been a reasonable local
choice, it may cause cascading errors later. Since we have the complete set of
minimum cost repair sequences available, we can select from that set a repair sequence
which causes fewer cascading errors. We thus
rank repair sequences by how far they allow parsing to continue successfully
(up to a threshold --- parsing the whole file would, in general, be too costly),
and choose from the subset that gets furthest (note that the time required to do
this is included in the 0.5s timeout). We thus also test a second hypothesis:

\begin{description}
  \item[H2] Ranking the complete set of minimum cost repair sequences by
    how far they allow parsing to continue locally reduces the global cascading
    error problem.
\end{description}

We also strongly validate this hypothesis. We do this by comparing `normal' \cpctplus
with a simple variant \cpctplusrev which reverses the ranking process, always selecting
from amongst the worst performing minimum cost repair
sequence. \cpctplusrev models the worst case of previous approaches in the
\fischer family, which non-deterministically select a single
minimum cost repair sequence. \cpctplusrev leads to \cpctplusreverrorlocsratioovercpctplus more
errors being reported (i.e.~it substantially worsens the cascading error problem).

This paper is structured as follows. We describe the \corchuelo algorithm
(Section~\ref{corchuelo}), filling in missing details from the original description
and correcting its definition. We then expand the algorithm into \cpctplus
(Section~\ref{corchueloplus}). Finally, we validate \cpctplus on a corpus of \corpussize real,
syntactically incorrect, Java programs comparing it to implementations
of panic mode and \corchuelo (Section~\ref{experiment}). To
emphasise that our algorithms are grammar-neutral, we show examples of
error recovery on different grammars in Appendix~\ref{app:examples}.

\section{Background}

We assume a high-level understanding of the mechanics of parsing in this paper,
but in this section we provide a handful of definitions, and a brief refresher
of relevant low-level details, needed to understand the rest of this paper.
Although the parsing tool we created for this paper is written in Rust, we
appreciate that this is still an unfamiliar language to many readers: algorithms
are therefore given in Python which, we hope, is familiar to most.

Although there are many flavours of parsing, the \fischer family
of error recovery algorithms are designed to be used with LR($k$)
parsers~\cite{knuth65lllr}. LR parsing remains one of the most widely used parsing
approaches due to the ubiquity of Yacc~\cite{johnson75yacc} and its
descendants (which include the Rust parsing tool we created for this paper).
We use Yacc syntax throughout this paper so that
examples can easily be tested in Yacc-compatible parsing tools.

Yacc-like tools take in a Context-Free Grammar (CFG) and produce a parser from
it. The CFG has one or more \emph{rules}; each rule has a name and one or more
\emph{productions} (often called `alternatives'); each production contains one
or more \emph{symbols}; and a symbol references either a \emph{token type}
or a grammar rule. One rule is designated the \emph{start rule}. The resulting parser
takes as input a stream of tokens, each of which has a \emph{type}
(e.g.~\texttt{INT}) and a \emph{value} (e.g.~\texttt{123}) -- we assume the
existence of a Lex-like tool which can split a string into a stream of tokens.
Strictly speaking, parsing is the act of
determining whether a stream of tokens is correct with respect to the underlying
grammar. Since this is rarely useful on its own, Yacc-like tools allow grammars
to specify `semantic actions' which are executed when a production in the grammar is
successfully matched. Except where stated otherwise, we assume that the semantic actions build
a \emph{parse tree}, ordering the tokens into a tree of nonterminal nodes
(which can have children) and terminal nodes (which cannot have children).

\begin{figure}[t]
\includegraphics{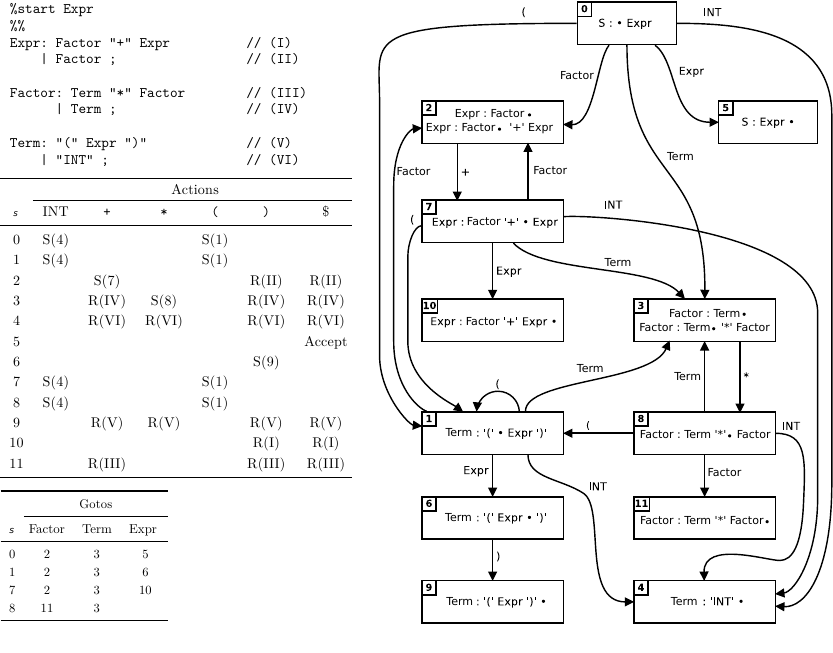}
\vspace{-23pt}
\caption{An example grammar (top left), its corresponding stategraph (right), and statetable
(split into separate action and goto tables; bottom left). Productions in
the grammar are labelled \texttt{(I)} to \texttt{(VI)}. In the stategraph: S($x$)
means `shift to state $x$'; R($x$) means `reduce production $x$ from the
grammar' (e.g.~\textit{action(3, `+')} returns R(IV) which references
the production `\texttt{Factor: Term;}').
  Each item within a state $[N \colon \alpha \bullet \beta]$ references one
  of rule $N$'s productions; $\alpha$ and $\beta$ each
represent zero or more symbols; with the \emph{dot} ($\bullet$) representing
  how much of the production must have been matched ($\alpha$) if parsing has
  reached that state, and how much remains ($\beta$).}
\label{fig:stategraphtable}
\label{fig:exprgrammar}
\end{figure}

The CFG is first transformed into a \emph{stategraph}, a statemachine
where each node contains one or more \emph{items} (describing the valid
parse states at that point) and edges are labelled with terminals or
nonterminals. Since even on a modern machine, a canonical
(i.e.~unmerged) LR stategraph can take several seconds to
build, and a surprising amount of memory to store, we use the
algorithm of \cite{pager77practical} to merge together compatible
states\footnote{\cite{pager77practical} can over-merge states when conflicts
occur \cite[p.~3]{denny10ielr} (i.e.~when Yacc uses
precedence rules to turn an ambiguous grammar into an unambiguous LR parser).
Since our error recovery approach operates purely on the statetable,
it should work correctly with other merging approaches such as that of~\cite{denny10ielr}.}. The
effect of this is significant, reducing the Java grammar we use later from
8908 to 1148 states. The stategraph is then transformed into a
\emph{statetable} with one row per state. Each row has a possibly empty \emph{action} (shift, reduce,
or accept) for each terminal and a possibly empty \emph{goto state} for each
nonterminal. Figure~\ref{fig:stategraphtable} shows an example grammar, its
stategraph, and statetable.

The statetable allows us to define a simple, efficient, parsing process. We
first define two functions relative to the statetable: \textsf{action}$(s, t)$
returns the action for the state $s$ and token $t$
or \emph{error} if no such action exists; and \textsf{goto}$(s, N)$
returns the goto state for the state $s$ and the nonterminal $N$ or \emph{error}
if no such goto state exists. We then define
a reduction relation $\lrarrow$ for $(\textit{parsing stack}, \textit{token
list})$ pairs with two reduction rules as shown in Figure~\ref{fig:lrreduction}.
A full LR parse $\lrarrowstar$ repeatedly applies the two $\lrarrow$ rules
until neither applies, which means that \textsf{action}$(s_n, t_0)$ is either:
$\textit{accept}$ (i.e.~the input has been fully parsed); or
$\textit{error}$ (i.e.~an error has been detected at the terminal $t_0$). A
full parse takes a starting pair of $([0], [t_0 \ldots t_n, \$])$,
where state $0$ is expected to represent the entry point into the stategraph, $t_0 \ldots t_n$
is the sequence of input tokens, and `$\$$' is the special End-Of-File (EOF) token.

\begin{figure}[t]
\centering
\begin{minipage}{0.63\textwidth}
\small
\[
\infer[\textsc{LR Shift}]
      {([s_0 \ldots s_n], [t_0 \ldots t_n]) \lrarrow ([s_0 \ldots s_n, s'], [t_1 \ldots t_n])}
      {\textsf{action}(s_n, t_0) = \textit{shift}\ s'}
\]
\end{minipage}
\vspace{-0pt}
\begin{minipage}{0.83\textwidth}
\[
\infer[\textsc{LR Reduce}]
      {([s_0 \ldots s_n], [t_0 \ldots t_n]) \lrarrow ([s_0 \ldots s_{n - \lvert \alpha \rvert}, s'], [t_0 \ldots t_n])}
      {(\textsf{action}(s_n, t_0) = \textit{reduce}\ N \colon \alpha)
       \wedge (\textsf{goto}(s_{n - \lvert \alpha \rvert}, N) = s')}
\]
\end{minipage}
\vspace{-0pt}
\caption{Reduction rules for $\lrarrow$, which operate on $(\textit{parsing stack},
\textit{token list})$ pairs. \textsc{LR Shift}
advances the input by one token and grows the parsing stack, while
\textsc{LR Reduce} unwinds (`reduces') the parsing stack when a production is
complete before moving to a new (`goto') state.}
\label{fig:lrreduction}
\end{figure}

\section{Panic mode}
\label{sec:panic mode}

Error recovery algorithms are invoked by a parser when it has yet to finish but
there is no apparent way to continue parsing (i.e.~when \textsf{action}$(s_n,
t_0) = \textit{error}$). Error recovery algorithms are thus called with a parsing
stack and a sequence of remaining input (which we represent
as a list of tokens): they can modify either or both of the
parsing stack and the input in their quest to get parsing back on track. The differences
between algorithms are thus in what modifications they can carry out
(e.g.~altering the parse stack; deleting input; inserting input), and how they
carry such modifications out.

The simplest grammar-neutral error recovery algorithms are widely called `panic mode'
algorithms (the origin of this family of algorithms seems lost in time).
While there are several members of this family for LL parsing,
there is only one fundamental way of creating
a grammar-neutral panic mode algorithm for LR parsing. We take our formulation from
\holub~\cite[p.~348]{holub90compilerdesign}\footnote{Note that step 2 in
\holub causes valid repairs to be missed: while it is
safe to ignore the top element of the parsing stack on the first iteration of
the algorithm, as soon as one token is skipped, one must check all elements of
the parsing stack. Our description simply drops step 2 entirely.}. The algorithm works by
popping elements off the parsing stack to
see if an earlier part of the stack is able to parse the next input symbol. If
no element in the stack is capable of parsing the next input symbol, the next input
symbol is skipped, the stack restored, and the process repeated. At worst,
this algorithm guarantees to find a match at the EOF token.
Figure~\ref{fig:holub:algorithm} shows a more formal version of this algorithm.

\begin{figure}[tb]
\begin{adjustbox}{valign=t,minipage=.53\textwidth}
\begin{lstlisting}[numbers=left,language=Python]
def holub(pstack, toks):
  while len(toks) > 0:
    npstack = pstack.copy()
    while len(npstack) > 0:
      if action(npstack[-1], toks[0]) != <!{\textrm{\textit{error}}}!>:
        return (npstack, toks)
      npstack.pop()
    del toks[0]
  return None
\end{lstlisting}
\end{adjustbox}
\hspace{5pt}
\begin{adjustbox}{valign=t,minipage=.45\textwidth}
\vspace*{-0pt}
\caption{Our version of the \holub~\cite{holub90compilerdesign} algorithm. This
panic mode algorithm takes in a (\emph{parsing stack}, \emph{token list})
pair and returns: a (\emph{parsing stack}, \emph{token list}) pair if
it managed to recover; or \texttt{None} if it failed to recover.
The algorithm tries to find an element in the stack that has a non-error action
for the next token in the input (lines 4--7). If it fails to find such an
element, the input is advanced by one element (line 8) and the stack restored
(line 3).}
\label{fig:holub:algorithm}
\end{adjustbox}
\vspace{-15pt}
\end{figure}

The advantage of this algorithm is its simplicity and speed. For example,
consider the grammar from Figure~\ref{fig:exprgrammar} and the input `\texttt{2
+ + 3}'. The parser encounters an error on the second `+' token, leaving it with
a parsing stack of [0, 2, 7] and the input `\texttt{+ 3}' remaining. The error
recovery algorithm now starts. It first tries \textsf{action}(7, `+') which
(by definition, since it is the place the parser encountered an error) returns
\textit{error}; it then pops the top element from the parsing stack and tries
\textsf{action}(2, `+'), which returns \textit{shift}. This is enough for the
error recovery algorithm to complete, and parsing resumes with a stack [0, 2].

The fundamental problem with error recovery can be seen from the above example:
by popping from the parsing stack, it implicitly deletes input from before the
error location (in this case the first `\texttt{+}') in order to find a way of
parsing input after the error location. This often leads to panic mode throwing
away huge portions of the input in its quest to find a repair. Not only can the
resulting recovery appear as a \emph{Deus ex machina}, but the more input
that is skipped, the more likely that a cascade of further parsing errors
ensues (as we will see later in Section~\ref{results}).

\section{\corchuelo}
\label{corchuelo}

There have been many attempts to create better LR error recovery algorithms than
panic mode. Most numerous are those error recovery algorithms in what we call the
\fischer family. Indeed, there are far too many
members of this family of algorithms to cover in one paper. We therefore start
with one of the most recent -- \corchuelo~\cite{corchuelo02repairing}. We first
explain the original algorithm (Section~\ref{corchuelo:orig}), although we use
different notation than the original, fill in several missing details, and
provide a more formal definition. We then
make two correctness fixes to ensure that the algorithm always
finds minimum cost repair sequences (Section~\ref{corchuelo:kimyi}). Since the original
gives few details as to how the algorithm might best be implemented, we then
explain our approach to making a fast implementation
(Section~\ref{corchuelo:implementation}).

\subsection{The original algorithm}
\label{corchuelo:orig}

Intuitively, the \corchuelo algorithm starts at the error state and tries
to find a minimum cost repair sequence consisting of: \textit{insert T}
(`insert a token of type T'), \textit{delete} (`delete the token at the current offset'),
or \textit{shift} (`parse the token at the current offset'). The
algorithm completes: successfully if it reaches an accept state or shifts
`enough' tokens ($N_\textit{shifts}$, set at 3 in \corchuelo);
or unsuccessfully if a repair sequence contains too many delete or insert repairs (set at 3 and 4
respectively in \corchuelo) or spans `too much' input ($N_\textit{total}$, set at 10
in \corchuelo). Repair sequences are reported back to users
with trailing \emph{shift} repairs pruned i.e.~[\emph{insert x, shift y, delete
z, shift a, shift b, shift c}] is reported as [\emph{insert x, shift y, delete
z}].

In order to find repair sequences, the algorithm keeps a breadth-first queue of
\emph{configurations}, each of which represents a different search state;
configurations are queried for their neighbours which are put into the queue;
and the search terminates when a successful configuration is found.
The cost of a configuration is the sum of the repair costs
in its repair sequence. By definition, a configuration's neighbours have the
same, or greater, cost to it.

\begin{figure}[tb]
\small
\centering
\begin{minipage}{0.9\textwidth}
\[
\infer[\textsc{CR Insert}]
      {([s_0 \ldots s_n], [t_0 \ldots t_n]) \crarrow ([s'_0 \ldots s'_m], [t_0 \ldots t_n], [\textit{insert}~t])}
      {\textsf{action}(s_n, t) \ne error \wedge t \ne \${}
       \wedge ([s_0 \ldots s_n], [t, t_0 \ldots t_n]) \lrarrowstar ([s'_0 \ldots s'_m], [t_0 \ldots t_n])}
\]
\end{minipage}
\vspace{-0pt}
\begin{minipage}{0.7\textwidth}
\[
\infer[\textsc{CR Delete}]
      {([s_0 \ldots s_n], [t_0, t_1 \ldots t_n])
       \crarrow
       ([s_0 \ldots s_n], [t_1 \ldots t_n], [\textit{delete}])}
      {t_0 \ne \$}
\]
\end{minipage}
\vspace{-0pt}
\begin{minipage}{0.75\textwidth}
\[
\hspace{-12pt}
\infer[\textsc{CR Shift 1}]
      {([s_0 \ldots s_n], [t_0 \ldots t_n])
       \crarrow
       ([s'_0 \ldots s'_m], [t_j \ldots t_n], [\underbrace{\textit{shift} \ldots \textit{shift}}_j])}
      {%
\begin{array}{c}
([s_0 \ldots s_n], [t_0 \ldots t_n]) \lrarrowstar ([s'_0 \ldots s'_m], [t_j \ldots t_n])
 \wedge 0 < j \leq N_\textit{shifts}
\\
 j = N_\textit{shifts} \vee
 \textsf{action}(s'_m, t_j) \in \{\textit{accept}, \textit{error}\} 
\end{array}}
\]
\end{minipage}
\vspace{-3pt}
\caption{The repair-creating reduction rules for \corchuelo.
\textsc{CR Insert} finds all terminals reachable from the current state and
creates insert repairs for them (other than the EOF token `$\$$').
\textsc{CR Delete} creates deletion repairs if user defined input remains.
\textsc{CR Shift 1} parses at least 1 and at most $N_\textit{shifts}$ tokens; if it reaches an accept or error
state, or parses exactly $N_\textit{shifts}$ tokens, then a shift repair per
token shifted is created.}
\label{fig:corchuelo:reductions}
\vspace*{-8pt}
\end{figure}

\begin{figure}[tb]
\begin{adjustbox}{valign=t,minipage=.52\textwidth}
\begin{lstlisting}[numbers=left,language=Python]
def corchueloetal(pstack, toks):
  todo = [[(pstack, toks, [])]]
  cur_cst = 0
  while cur_cst < len(todo):
    if len(todo[cur_cst]) == 0:
      cur_cst += 1
      continue
    n = todo[cur_cst].pop()
    if action(n[0][-1], n[1][0]) == <!{\textrm{\textit{accept}}}!> \
       or ends_in_N_shifts(n[2]):
      return n
    elif len(n[1]) - len(toks) == N_total:
      continue
    for nbr in all_cr_star(n[0], n[1]):
      if len(n[2]) > 0 and n[2][-1] == <!{\textrm{\textit{delete}}}!> \
         and nbr[2][-1] == <!{\textrm{\textit{insert}}}!>:
        continue
      cst = cur_cst + rprs_cst(nbr[2])
      for _ in range(len(todo), cst):
        todo.push([])
      todo[cst].append((nbr[0], nbr[1], \
                        n[2] + nbr[2]))
  return None

def rprs_cst(rprs):
  c = 0
  for r in rprs:
    if r == <!{\textrm{\textit{shift}}}!>: continue
    c += 1
  return c

def all_cr_star(pstack, toks):
  # Exhaustively apply the <!{$\crarrowstar$}!> relation to
  # (pstack, toks) and return the resulting
  # list of (pstack, toks, repair) triples.
\end{lstlisting}
\end{adjustbox}
\hspace{5pt}
\begin{adjustbox}{valign=t,minipage=.46\textwidth}
\vspace*{-0pt}
\caption{Our version of the \corchuelo algorithm. The main function
\texttt{corchueloetal} takes in a (\emph{parsing stack}, \emph{token list})
pair and returns: a (\emph{parsing stack}, \emph{token list}, \emph{repair
sequence}) triple where \emph{repair sequence} is guaranteed to be a minimum
cost repair sequence; or \texttt{None} if it failed to find a repair sequence.\\[9pt]
The algorithm maintains a todo list of lists: the first sub-list contains
configurations of cost 0, the second sub-list configurations of cost 1, and
so on. The todo list is initialised with the error parsing stack, remaining
tokens, and an empty repair sequence (line 2). If there are todo items left, a
lowest cost configuration $n$ is picked (lines 4--8). If $n$ represents an accept state (line
9) or if the last $N_\textit{shifts}$ repairs are shifts (line 10), then $n$
represents a minimum cost repair sequence and the algorithm terminates
successfully (line 11). If $n$ has already consumed $N_\textit{total}$ tokens,
then it is discarded (lines 12, 13). Otherwise, $n$'s neighbours
are gathered using the $\crarrow$ relation (lines 14, 32--35). To avoid
duplicate repairs, \textit{delete} repairs never follow \textit{insert} repairs
(lines 15--17). Each neighbour has its repairs costed (line 18) and is then
assigned to the correct todo sub-list (lines 21--22).\\[9pt]
The \texttt{rprs\_cst} function returns the cost of a repair sequence. Inserts
and deletes cost 1, shifts 0.}
\label{fig:corchuelo:algorithm}
\end{adjustbox}
\vspace{-15pt}
\end{figure}

As with the original, we explain the approach in two parts.
First is a new reduction relation $\crarrow$ which defines a configuration's
neighbours (Figure~\ref{fig:corchuelo:reductions}). Second is an algorithm which
makes use of the $\crarrow$ relation to generate neighbours, and determines when
a successful configuration has been found or if error recovery has failed
(Figure~\ref{fig:corchuelo:algorithm}).
As well as several changes for clarity, the biggest difference is that
Figure~\ref{fig:corchuelo:algorithm} captures semi-formally what
\corchuelo explain in prose (spread amongst
several topics over several pages): perhaps inevitably
we have had to fill in several missing details. For example,
\corchuelo do not define what the cost of repairs is: for
simplicities sake, we define the cost of \textit{insert} and \textit{delete} as
1, and \textit{shift} as 0.\footnote{It is trivial to extend this to variable
token costs if desired, and our implementation supports this. However, it is
unclear whether non-uniform token costs are useful in practise
\cite[p.96]{cerecke03phd}.}

\subsection{Ensuring that minimum cost repair sequences aren't missed}
\label{corchuelo:kimyi}

\textsc{CR Shift 1} (see Figure~\ref{fig:corchuelo:reductions}) has two flaws
which prevent it from generating all possible minimum cost repair sequences.

\begin{figure}[tb]
\small
\[
\hspace{-5pt}
\infer[\textsc{CR Shift 2}]
      {([s_0 \ldots s_n], [t_0 \ldots t_n]) \crarrow ([s'_0 \ldots s'_m], [t_j \ldots t_n], [\underbrace{\textit{shift} \ldots \textit{shift}}_j])}
      {%
\begin{array}{c}
([s_0 \ldots s_n], [t_0 \ldots t_n]) \lrarrowstar ([s'_0 \ldots s'_m], [t_j \ldots t_n])
 \wedge 0 \leq j \leq N_\textit{shifts}
\\
 (j = 0 \wedge [s_0 \ldots s_n] \ne [s'_0 \ldots s'_m]) \vee j = N_\textit{shifts} \vee
 \textsf{action}(s'_m, t_j) \in \{\textit{accept}, \textit{error}\}
\end{array}}
\]

\[
\hspace{20pt}
\infer[\textsc{CR Shift 3}]
      {([s_0 \ldots s_n], [t_0 \ldots t_n]) \crarrow ([s'_0 \ldots s'_m], [t_j \ldots t_n], R)}
      {%
\begin{array}{c}
([s_0 \ldots s_n], [t_0 \ldots t_n]) \lrarrowstar ([s'_0 \ldots s_m], [t_j \ldots t_n])
 \wedge 0 \leq j \leq 1
\\
 (j = 0 \wedge [s_0 \ldots s_n] \ne [s'_0 \ldots s_m] \wedge R = [])
  \vee
  (j = 1 \wedge R = [\textit{shift}])
\end{array}
}
\]
\vspace{-15pt}
\caption{\textsc{CR Shift 1} always consumes input, when sometimes performing
one or more reduction/gotos without consuming input would be better. \textsc{CR
Shift 2} addresses this issue. Both \textsc{CR Shift 1} and \textsc{CR Shift 2}
generate multiple shift repairs in one go, which causes them to skip
`intermediate' (and sometimes important) configurations. \textsc{CR Shift 3}
generates at most one shift, exploring all intermediate configurations.}
\label{fig:corchuelo:kimyi}
\vspace*{-0pt}
\end{figure}

\begin{figure}[t]
\hspace*{0pt}
\begin{tabular}{llll}
\begin{subfigure}{0.012\textwidth}
\vspace{-22pt}
\caption{}
\label{lst:crshift2:crshift2}
\end{subfigure}
&
\begin{minipage}{190pt}
\begin{lstlisting}
Delete 3, Delete +
Delete 3, Shift +, Insert Int
Insert +, Shift 3, Shift +, Insert Int
Insert *, Shift 3, Shift +, Insert Int
\end{lstlisting}
\end{minipage}
&
\begin{subfigure}{0.012\textwidth}
\vspace{-22pt}
\caption{}
\label{lst:crshift2:crshift3}
\end{subfigure}
&
\begin{minipage}{140pt}
\vspace{-16pt}
\begin{lstlisting}
Insert *, Shift 3, Delete +
Insert +, Shift 3, Delete +
\end{lstlisting}
\end{minipage}
\end{tabular}
\vspace{-5pt}
\caption{Given the input `\texttt{2 3 +}' and the grammar from
Figure~\ref{fig:exprgrammar}, \textsc{CR Shift 1} is unable to find any repair
sequences because it does not perform the reductions/gotos necessary after
the final \textit{insert} or \textit{delete} repairs to reach an accept state.
(\subref{lst:crshift2:crshift2}) \textsc{CR Shift 2} can find 4 minimum cost repair sequences.
(\subref{lst:crshift2:crshift3}) \textsc{CR Shift 3} can find a further 2
minimum cost repair sequences on top those found by \textsc{CR Shift 2} (i.e.~6 in total).}
\label{fig:crshift2:example}
\end{figure}

First, \textsc{CR Shift 1} always consumes input, missing intermediate
configurations (including \textit{accept} states!) that only require
reductions/gotos to be performed. \textsc{CR Shift 2} in Figure~\ref{fig:corchuelo:kimyi}
shows the two-phase fix which addresses this problem.
We first change the condition $0 < j \leq N_\textit{shifts}$ to $0 \leq j \leq
N_\textit{shifts}$ so that the parser can make progress without consuming
input. However, this opens the possibility of an infinite loop, so
we then add a condition that if no input is consumed, the parsing stack
must have changed. In other words, we require progress
to be made, whether or not that progress involved consuming input.

Second, \textsc{CR Shift 1} and \textsc{CR Shift 2} generate multiple shifts
at a time. This causes them to skip intermediate
configurations from which minimum cost repair sequences may be found.
The solution\footnote{The problem, and the basis of a fix, derive from
\cite[p.~12]{kimyi10astar}, though their suggestion suffers from the same
problem as \textsc{CR Shift 1}.} is simple: at most one shift can be generated at any one
time. \textsc{CR Shift 3} in
Figure \ref{fig:corchuelo:kimyi} (as well as incorporating the fix from
\textsc{CR Shift 2}) generates at most one shift repair at a time. Relative to
\textsc{CR Shift 1}, it is simpler, though it also inevitably slows down the
search, as more configurations are generated.

The problems with \textsc{CR Shift 1}, in particular, can be severe.
Figure~\ref{fig:crshift2:example} shows an example input where \textsc{CR Shift
1} is unable to find any repair sequences, \textsc{CR Shift 2} some, and
\textsc{CR Shift 3} all minimum cost repair sequences.

\subsection{Implementation considerations}
\label{corchuelo:implementation}

\label{cactusconfigurations}
The definitions we have given thus far do not obviously lead to
an efficient implementation and \corchuelo give few useful
hints. We found that two techniques were both
effective at improving performance while being simple to implement.

First, \corchuelo suggest using a breadth-first search but give no
further details. It was clear to us that the most natural way to model the search is as an instance of
Dijkstra's algorithm. However, rather than use a general queue data-structure (probably based on a
tree) to discover which element to search next, we use a similar queue data-structure to
\cite[p.~25]{cerecke03phd}. This consists of one sub-list per cost (i.e.~the
first sub-list contains configurations
of cost 0, the second sub-list configurations of cost 1 and so on).
Since we always know what cost we are currently investigating,
finding the next todo element requires only a single \texttt{pop}
(line 8 of Figure~\ref{fig:corchuelo:algorithm}). Similarly,
adding elements requires only an \texttt{append} to the relevant sub-list
(lines 18, 21, 22). This
data-structure is a good fit because costs in our setting are always small
(double digits is unusual for real-world
grammars) and each neighbour generated from a configuration with cost $c$ has
a cost $\geq c$.

Second, since error recovery frequently adjusts and resets parsing stacks and
repair sequences, during error recovery we do not represent these as lists (as
is the case during normal parsing). We found that lists consume noticeably more
memory, and are slightly less
efficient, than using parent pointer trees (often called `cactuses').
Every node in such a tree has a reference to a single parent (or \texttt{null} for the
root node) but no references to child nodes. Since our implementation is written
in Rust -- a language without garbage collection -- nodes
are reference counted (i.e.~a parent is only freed when it is not in a todo list and
no children point to it). When the error recovery algorithm starts, it
converts the main parsing stack (a list) into a parent pointer tree; and
repair sequences start as empty parent pointer trees. The $\crarrow$ part
of our implementation thus operates exclusively on parent pointer trees.
Although this does mean that neighbouring configurations are scattered throughout
memory, the memory sharing involved seems to more than compensate for
poor cache behaviour; it also seems to be a good
fit with modern \texttt{malloc} implementations, which are particularly
efficient when allocating and freeing objects of the same size.
This representation is likely to be a reasonable choice in most contexts, although it is
difficult to know from our experience whether it will always be the best choice (e.g~for
garbage collected languages).

One seemingly obvious improvement is to split the search into parallel threads. However,
we found that the nature of the problem means that parallelisation is more
tricky, and less productive, than might be expected. There are two related
problems: we cannot tell in advance if a given
configuration will have huge numbers of successors or none at all; and
configurations are, in general, searched for successors extremely quickly. Thus
if we attempt to seed threads with initial sets of configurations, some threads
quickly run out of work whilst others have ever growing queues. If,
alternatively, we have a single global queue then significant amounts of time
can be spent adding or removing configurations in a thread-safe manner.
A work stealing algorithm might solve this problem but, as we shall see in
Section~\ref{experiment}, \cpctplus runs fast enough that the additional
complexity of such an approach is not, in our
opinion, justified.

\section{\cpctplus}
\label{corchueloplus}

In this section, we extend the \corchuelo algorithm to become what we call
\cpctplus. First we enhance the algorithm to find
the complete set of minimum cost repair sequences
(Section~\ref{corchuelo:allminimumcost}). Since this slows down
the search, we optimise by merging together compatible configurations
(Section~\ref{merging}). The complete set of
minimum cost repair sequences allows us to make an
algorithm less susceptible to the cascading error problem
(Section~\ref{rankingrepairs}). We then change the criteria for terminating
error recovery (Section~\ref{timeout}).

\subsection{Finding the complete set of minimum cost repair sequences}
\label{corchuelo:allminimumcost}

The basic \corchuelo algorithm non-deterministically completes as soon
as it has found a single minimum cost repair sequence. This is confusing
in two different ways: the successful repair sequence found can vary from run
to run; and the successful repair sequence might not match the user's intention.

We therefore introduce the idea of the complete set of minimum cost repair sequences: that
is, all equivalently good repair sequences. Although we will refine the concept
of `equivalently good' in Section~\ref{rankingrepairs}, at this stage we
consider all successful repair sequences with minimum cost $c$ to be
equivalently good. In other words, as soon as we find the first successful repair
sequence, its cost $c$ defines the minimum cost.

An algorithm to generate this set is then simple:
when a repair sequence of cost $c$ is found to
be successful, we discard all repair sequences with cost $> c$, and continue
exploring configurations in cost $c$ (including, transitively, all neighbours that are
also of cost $c$; those with cost $> c$ are immediately discarded). Each
successful configuration is recorded and, when all configurations
in $c$ have been explored, the set of successful configurations is returned.
One of these successful configurations is then non-deterministically chosen,
applied to the input, and parsing continued.

\subsection{Merging compatible configurations}
\label{merging}

Relative to finding a single solution, finding the complete set of repair
sequences can be extremely expensive because there may
be many remaining configurations in $c$, which may, transitively, have many neighbours.
Our solution to this performance problem is to merge together \emph{compatible}
configurations on-the-fly, preserving their distinct repair sequences while
reducing the search space. Two configurations are compatible if:

\begin{enumerate}
\item their
parsing stacks are identical,
\item they both have an identical amount of input remaining,
\item  and their repair sequences are compatible.
\end{enumerate}

\noindent Two repair sequences are compatible:

\begin{enumerate}
   \item if they both end in the same number ($n \ge 0$) of shifts,
   \item and, if one repair sequence ends in a delete, the other repair sequence also
ends in a delete.
\end{enumerate}

\noindent The first of these conditions is a direct consequence of the fact that a configuration
is deemed successful if it ends in $N_\textit{shifts}$ shift
repairs. When we merge configurations, one part of the merge is `dominant'
(i.e.~checked for $N_\textit{shifts}$) and the other `subsumed': we have to
maintain symmetry between the two to prevent the
dominant part accidentally preventing the subsumed part from being recorded as
successful. In other words, if the dominant part of the merge had fewer shifts
at the end of its repair sequence than the subsumed part, then the
$N_\textit{shifts}$ check (line 10, Figure~\ref{fig:corchuelo:algorithm}) would
fail, even though reversing the dominant and subsumed
parts may have lead to success. It is therefore only safe to merge repair sequences
which end in the same number of shifts.

The second condition relates to the weak form of compatible merging inherited
from \cite[p.~8]{corchuelo02repairing}: delete repairs are never followed by an
insert (see Figure~\ref{fig:corchuelo:algorithm}) since [\textit{delete},
\textit{insert x}] always leads to the same configuration as [\textit{insert x},
\textit{delete}]. Although we get much of the same effect through
compatible configuration merging, we keep it as a separate optimisation because: it is
such a frequent case; our
use of the todo list means that we would not catch every case; the
duplicate repair sequences are uninteresting from a user perspective, so we
would have to filter them out later anyway; and each additional merge costs
memory. We thus have to make sure that merged repair sequences don't accidentally
suppress insert repairs because one part of the repair sequence ends in a delete
while the other does not. The simplest way of solving this problem is thus to
forbid merging repair sequences if one sequence ends in a delete and the other does not.

Fortunately, implementing compatible configuration merging is simple. We
first modify the todo data-structure to be a
list-of-ordered-hashsets\footnote{An ordered hashset preserves insertion order,
and thus allows list-like integer indexing as well as hash-based lookups.}. This has
near-identical \texttt{append} / \texttt{pop} performance to a normal list, but
filters out duplicates with near-identical performance to an unordered hashset.
We then make use of a simple property of hashsets: an object's hashing behaviour
need only be a non-strict subset of its equality behaviour. Configuration
hashing is based solely on a configuration's parsing stack and remaining input,
giving us a fast way of finding configurations that are compatible under
conditions \#1 (identical parsing stacks) and \#2 (identical input remaining).
As well as checking those two conditions, configuration equality also checks
condition \#3 (compatible repair sequences).

Conceptually, merging two configurations together is simple: each configuration
needs to store a set of repair sequences, each of which is updated as further
repairs are found. However, this is an extremely inefficient representation as
the sets involved need to be copied and extended as each new repair is found.
Instead, we reuse the idea of graph-structured stacks from GLR
parsing~\cite[p.~4]{tomita87efficient} which allows us to avoid copying whenever
possible. The basic idea is that configurations no longer reference a parent pointer tree of
repairs directly, but instead a parent pointer tree of \emph{repair merges}. A
repair merge is a pair (\textit{repair}, \textit{merged}) where
\textit{repair} is a plain repair and \textit{merged} is a (possibly null) set of
repair merge sequences. This structure has two advantages. First, the
$N_\textit{shifts}$ check can be performed solely using the first element of
repair merge pairs. Second, we avoid allocating memory for configurations which
have not yet been subject to a merge. The small
downside to this scheme is that expanding configurations into repair sequences requires
recursively expanding both the normal parent pointer tree of the first
element as well as the merged parent pointer trees of the second element.

Compatible configuration merging is particularly effective in complex cases,
even though it can only merge configurations in the todo list (i.e.~we cannot detect
all possible compatible merges). An example of compatible configuration merging
can be seen in Figure~\ref{fig:cpctplus:full}.

\subsection{Ranking repair sequences}
\label{rankingrepairs}

\begin{figure}[t]
\includegraphics[width=1.0\textwidth]{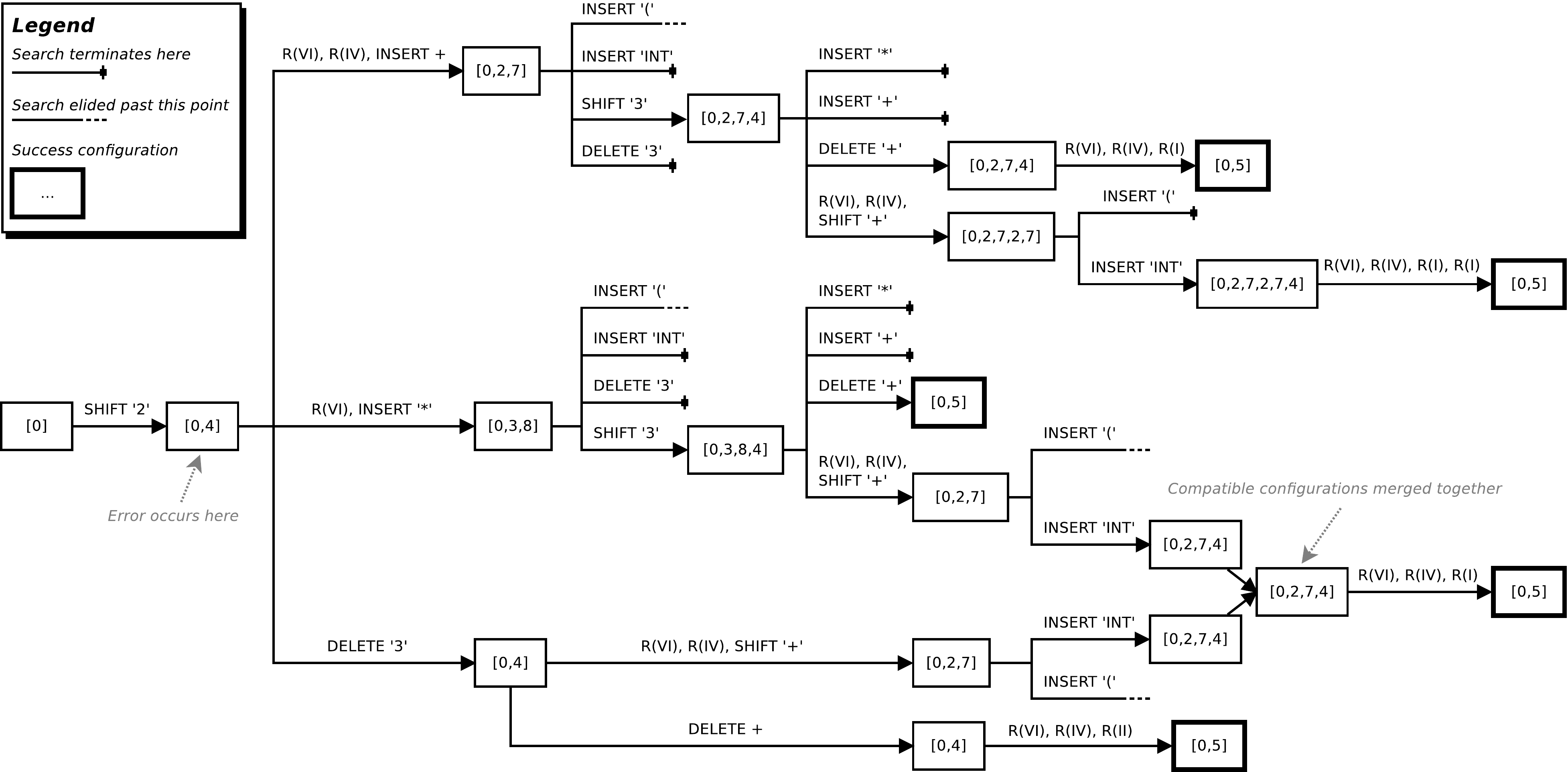}
\caption{An elided visualisation of a real run of \cpctplus with the input
`2 3 +' and the grammar from Figure~\ref{fig:exprgrammar}. The left hand side of
the tree shows the `normal' parser at work, which hits an error as soon as it
has shifted the token `\texttt{2}': at this point, \cpctplus starts operating.
As this shows, the search encounters various dead ends, as well as successful
routes. As shown in Figure~\ref{fig:crshift2:example}, this input has 6 minimum
cost repair sequences, but the search only has 5 success configurations, because
two configurations were merged together.}
\label{fig:cpctplus:full}
\end{figure}

\begin{figure}[t]
\hspace{5pt}
\begin{tabular}{p{0.02\textwidth}p{0.35\textwidth}p{0.02\textwidth}p{0.55\textwidth}}
\begin{subfigure}{0.02\textwidth}
\caption{}
\label{lst:ranking:java}
\end{subfigure}
&
\begin{minipage}[t]{0.35\textwidth}
\vspace{-14pt}
\begin{lstlisting}[language=Java]
class C {
    T x = 2 +
    T y = 3;
}
\end{lstlisting}
\end{minipage}
&
\begin{subfigure}{0.02\textwidth}
\caption{}
\label{lst:ranking:output}
\end{subfigure}
&
\begin{minipage}[t]{0.45\textwidth}
\vspace{-14pt}
\begin{lstlisting}
Parsing Error at line 3 col 7. Repair
sequences found:
  1: Insert ,
\end{lstlisting}
\end{minipage}
\end{tabular}
\vspace{-10pt}
\caption{An example showing how the ranking of repair sequences can lessen the
cascading error problem. The Java example (\subref{lst:ranking:java}) leads
to a parsing error on line 3 at `\texttt{y}', with three minimum cost repair
sequences found: [\textit{insert} \texttt{,}], [\textit{insert} \texttt{?}], and
[\textit{insert} \texttt{(}]. These repair sequences are then ranked by how far
they allow parsing to continue successfully. [\textit{insert} \texttt{,}] leads
to the rest of the file being parsed without further error. [\textit{insert}
\texttt{?}] causes a cascading error at `\texttt{;}' which must then be resolved
by completing the ternary expression started by `?' (e.g.~changing line 3 to
`\texttt{T ? y : this;}'). Similarly, [\textit{insert} \texttt{(}] causes a
cascading error at `\texttt{;}' which must then be resolved by inserting a
`\texttt{)}'. Since [\textit{insert} \texttt{,}] is ranked more highly than the
other repair sequences, the latter are discarded, leading to the parsing output shown
in (\subref{lst:ranking:output}). javac in contrast attempts to insert
`\texttt{;}' before `\texttt{y}' causing a cascading error on the next token.}
\label{fig:ranking}
\end{figure}

In nearly all cases, members of the complete set of minimum cost repair
sequences end with $N_\textit{shifts}$ (the only exception being error
locations near the end of an input where recovery leads to an accept state). Thus
while the repair sequences we find are all equivalently good within the range
of $N_\textit{shifts}$, some, but not others, may perform poorly beyond that
range. This problem is exacerbated by the fact that $N_\textit{shifts}$ has
to be a fairly small integer (we use 3, the value suggested by \corchuelo)
since each additional token searched exponentially increases the
search space. Thus while all repair sequences found may be locally
equivalent, when considered in the context of the entire input, some may be
better than others. While it is, in general, impractically slow to determine
which repair sequences are the global best, we can quickly determine which are
better under a wider definition of `local'.

We thus rank configurations which represent the
complete set of minimum cost repair sequences by how far they allow parsing to continue, up to a limit of
$N_\textit{try}$ tokens (which we somewhat arbitrarily set at 250). Taking
the furthest-parse point as our top rank, we then discard all configurations
which parsed less input than this. The reason
why we rank the configurations, and not the repair sequences, is that we only
need to rank one of the repair sequences from each merged configuration, a small but useful
optimisation. We then expand the top ranked configurations into repair
sequences and remove shifts from the end of those repair sequences. Since the
earlier merging of compatible configurations is imprecise (it misses
configurations that have already been processed), there can be some remaining
duplicate repair sequences: we thus perform a final purge of duplicate repair
sequences. Figure~\ref{fig:cpctplus:full} shows a visualisation of \cpctplus in
action.

Particularly on real-world grammars, selecting the top-ranked repair sequences
substantially decreases cascading errors (see Figure~\ref{fig:ranking} for an example).
It also does so for very little additional computational cost, as the complete set of
minimum cost repair sequences typically contains only a small number of items.
However, it cannot entirely reduce the cascading error problem. Since,
from our perspective, each member of the top-ranked set is equivalently good, we
non-deterministically select one of its members to repair the input and allow
parsing to continue. This can mean that we select a repair sequence which
performs less well beyond $N_\textit{try}$ tokens than other repair sequences in
the top-ranked set.

\subsection{Timeout}
\label{timeout}

The final part of \cpctplus relates to the use of $N_\textit{total}$ in
\corchuelo. As with all members of the
\fischer family, \cpctplus is not only unbounded in
time~\cite[p.~14]{mckenzie95error}, but
also unbounded in memory. In an attempt to combat this, \corchuelo limits
repair sequences to a maximum of 3 deletes and 4 inserts and a span of at most
10 tokens, attempting to stop the search from going too far. Unfortunately it
is impossible to find good values for these constants, as `too far' is entirely
dependent on the grammar and erroneous input: Java's grammar, for example, is
large with a commensurately large search space (requiring smaller constants)
while Lua's grammar is small with a commensurately small search space (which
can cope with larger constants).

This problem can be easily seen on inputs with unbalanced brackets
(e.g.~expressions such as `\texttt{x = f(();}'): each additional unmatched
bracket exponentially increases the search space. On a modern machine
with a Java 7 grammar, \cpctplus takes about 0.3s to find
the complete set of minimum cost repair sequences for 3 unmatched brackets, 3s
for 4 unmatched brackets, and 6 unmatched brackets caused our 32GiB test machine
to run out of RAM.

The only sensible alternative is a timeout: up to several seconds is safe in our
experience. We thus remove $N_\textit{total}$ from \cpctplus and rely
entirely on a timeout which, in this paper, is defined to be 0.5s.

\section{Experiment}
\label{experiment}

In order to understand the performance of \cpctplus, we conducted a large
experiment on real-world Java code. In this section we outline our methodology
(Section~\ref{methodology}) and results (Section~\ref{results}). Our experiment
is fully repeatable and downloadable from
\url{https://archive.org/download/error_recovery_experiment/0.4/}. The results
from our particular run of the experiment can also be downloaded from the same
location.

\subsection{Methodology}
\label{methodology}

In order to evaluate error recovery implementations, we need a concrete implementation. We
created a new Yacc-compatible parsing system \emph{grmtools} in Rust which
we use for our experiments. Including
associated libraries for LR table generation and so on, \emph{grmtools} is around
13KLoC. Although intended as a production library, it has accidentally
played a part as a flexible test bed for experimenting with, and understanding,
error recovery algorithms. We added a simple front-end \emph{nimbleparse} which produces the output
seen in e.g.~Figure~\ref{fig:javaerror}.

There are two standard problems when evaluating error recovery algorithms: how
to determine if a good job has been done on an individual example; and how to obtain
sufficient examples to get a wide perspective on an algorithm's performance. Unfortunately,
solutions to these problems are mutually exclusive, since the only way to tell if
a good job has been done on a particular is to
manually evaluate it, which means that it is only practical to use a small set
of input programs. Most papers we are aware of use at most 200 source files
(e.g.~\cite{corchuelo02repairing}), with one using a single source file with minor
variants~\cite{kimyi10astar}. \cite{cerecke03phd} was the first to use a
large-scale corpus of approximately 60,000 Java source files. Early in the
development of our methodology, we performed some rough experiments which
suggested that statistics only start to stabilise once a corpus exceeds 10,000
source files. We therefore prefer to use a much larger corpus than most previous
studies. We are fortunate to have access to the Blackbox
project~\cite{brown14blackbox}, an opt-in data collection facility for the BlueJ
editor, which records major editing events (e.g.~compiling a file) and sends
them to a central repository. Crucially, one can see the source code associated
with each event. What makes Blackbox most appealing as a data source is its
scale and diversity: it has hundreds of thousands of users, and a huge
collection of source code.

We first obtained a Java 1.5 Yacc grammar and updated it to support Java
1.7.\footnote{Unfortunately, changes to the method calling syntax in Java 1.8
mean that it is an awkward, though not impossible, fit for an LR(1) formalism
such as Yacc, requiring substantial changes to the current Java Yacc grammar. We
consider the work involved beyond that useful for this paper.} We then
randomly selected source files from Blackbox's database (following the lead of
\cite{santos18syntax}, we selected data from Blackbox's beginning until the end
of 2017-12-31). We then ran such source files through our Java 1.7 lexer. We immediately
rejected files which didn't lex, to ensure we were dealing solely
with parsing errors\footnote{Happily, this also excludes files which can't possibly be
Java source code. Some odd things are pasted into text
editors.} (see Section~\ref{sec:lexing errors into parsing
errors}). We then parsed candidate files with our Java grammar and rejected any
which did parse successfully, since there is little point running an error
recovery algorithm on correct input. The final corpus consists of \corpussize source files
(collectively a total of \corpussizemb{}MiB). Since Blackbox, quite reasonably,
requires each person with access to the source files to register with them, we
cannot distribute the source files directly; instead, we distribute the
(inherently anonymised) identifiers necessary to extract the source
files for those who register with Blackbox.

The size of our corpus means that we cannot manually evaluate repairs for
quality. Instead, we report several other metrics, of which the number of error
locations is perhaps the closest proxy for perceived quality. However,
this number has to be treated with caution for two reasons. First, it is
affected by differences in the failure rate: if a particular error recovery
algorithm cannot repair an entire file then it may not have had time to find all the
`true' error locations. Second, the number of error locations only allows
relative comparisons. Although we know that the corpus
contains at least \corpussize manually created errors (i.e.~at least one per
file), we cannot know if, or how many, files contain more than one error.
Since we cannot know the true number of error locations, we are unable
to evaluate algorithms in an absolute sense.

In order to test hypothesis H1 we ran each error recovery algorithm against
the entire Java corpus, collecting for each file: the time spent in recovery (in seconds);
whether error recovery on the file succeeded or failed (where failure is due
to either the timeout being exceeded or no repair sequences being found for
an error location); the
number of error locations; the cost of repair sequences at each
error location;
and the proportion of tokens skipped by error recovery (i.e.~how many \emph{delete}
repairs were applied).
We measure the time spent in error recovery with a monotonic wall-clock timer,
covering the time from when the main parser first invokes
error recovery until an updated parsing stack and parsing index are returned
along with minimum cost repair sequences. The timer is suspended when normal parsing
restarts and resumed if error recovery is needed again (i.e.~the timeout applies
to the file as a whole).

We evaluate three main error recovery algorithms: \corchuelo, \cpctplus, and
panic mode. Our implementation of \corchuelo is to some extent a
`best effort' since we have had to fill in several implementation details ourselves. As
per the description, we: use the same
limits on repair sequences (repair sequences can contain at most 3
delete or 4 insert repairs, and cannot span more than 10 tokens in the input);
complete as soon as a single successful repair sequence is found; and, when no
available repair sequence is found, fall back on panic mode. In addition, we
impose the same 0.5s timeout on this algorithm, as it is otherwise unbounded in
length, and sometimes exhausts available RAM. Panic mode implements the algorithm
from Section~\ref{sec:panic mode}. We do not report the
average cost size for panic mode or \corchuelo
(which falls back on panic mode) since they do not (always) report
repair sequences (see Section~\ref{sec:panic mode}). Although
panic mode can implicitly delete input from before the error location, we
only include the input it explicitly skips in the proportion of tokens skipped.

In order to test hypothesis H2, we created a variant of \cpctplus called \cpctplusrev.
Instead of selecting from the minimum cost repair sequences which allow
parsing to continue furthest, \cpctplusrev selects from those which allow parsing to
continue the least far. This models the worst case for other members of the
\fischer family which non-deterministically select a single minimum
cost repair sequence. In other words, it allows us to understand how many
more errors could be reported to users of other members of the
\fischer family compared to \cpctplus.

Configuration merging (see Section~\ref{merging}) is the most complex part of
\cpctplus. To understand whether this complexity leads to better performance, we
created another variant of \cpctplus called \cpctplusdontmerge
which disables configuration merging.

We bootstrap~\cite{efron79bootstrap} our results
\numbootstrap times to produce 99\% confidence intervals. However, as
Figure~\ref{fig:results:cpctplus_histogram} shows, our distribution is heavy-tailed,
so we cannot bootstrap naively. Instead, we ran each error recovery algorithm
\numruns times on each source file; when bootstrapping we randomly sample one of the
\numruns values collected (i.e.~our bootstrapped data contains an entry for every
file in the experiment; that entry is one of the \numruns values collected for
that file).

All experiments were run on an otherwise unloaded Intel Xeon E3-1240 v6 with
32GiB RAM running Debian 10. We disabled hyperthreading and turbo boost and ran experiments
serially. Our experiments took approximately 15 days to complete. We used Rust
1.43.1 to compile \emph{grmtools} (the
\texttt{Cargo.lock} file necessary to reproduce the build is included in our
experimental repository).

\subsection{Results}
\label{results}

\begin{figure}[t]
\hspace{-9pt}
\begin{tabular}{lcccccc}
\toprule
  & Mean     & Median   & Cost      & Failure   & Tokens      & Error \\
  & time (s) & time (s) & size (\#) & rate (\%) & skipped (\%) & locations (\#) \\
\midrule
\input{table.tex}
\bottomrule
\end{tabular}
\caption{Summary statistics from running various error recovery algorithms over
a corpus of \corpussize Java files (for all measures, lower is better).
Mean and median times
report how much time was spent in error recovery per file: both figures
include files which exceeded the recovery timeout, so they represent the `real'
times that users would experience, whether or not all errors are repaired or
not. Cost size reports the mean cost (i.e.~the number of insert and delete
repairs) of each error location repaired (this number is meaningless for
\corchuelo -- which falls back on panic mode -- and panic mode, since panic mode does
not report repair sequences).
The failure rate is the percentage of files which could not be fully
repaired within the timeout. Tokens skipped is
the proportion of input
skipped (because of a delete repair). \cpctplusrev models the worst case
of non-deterministic \fischer algorithms by reversing the order of repair
ranking (see Section~\ref{rankingrepairs}). \cpctplusdontmerge shows the
performance of \cpctplus if configuration merging is disabled (see
Section~\ref{merging}).}
\label{fig:results:summary}
\end{figure}

\begin{figure}[t]
\vspace{-10pt}
\includegraphics[scale=.7]{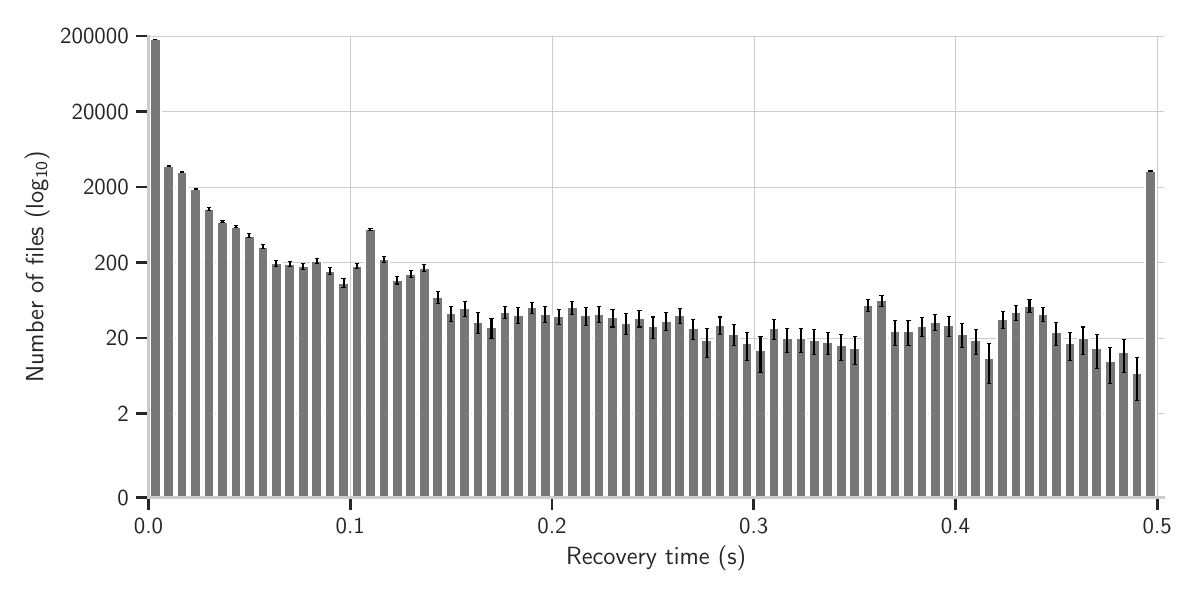}
\vspace{-25pt}
\caption{A histogram of the time spent in error recovery by \cpctplus for files in our
corpus. The $x$ axis shows time (up to the timeout of 0.5s) and the $y$ axis is
a logarithmic scale for the number of files. Error bars represent 99\% confidence
intervals. As this clearly shows, the vast majority of files fit in the
histogram's first bin; there is then a gradual decrease until around 0.15s,
with a broadly flat distribution from then until the pronounced peak
at the timeout of 0.5s. Figure~\ref{fig:results:cpctplus_longer_histogram}
in the Appendix shows how extending the timeout increases the number of files
which can be successfully recovered.}
\label{fig:results:cpctplus_histogram}
\end{figure}

Figure~\ref{fig:results:summary} shows a summary of the results of our
experiment. Comparing the different algorithms requires care as a
higher failure rate tends to lower the cost size, tokens skipped, and number of
error locations simply because files are not completely parsed. For example,
although \corchuelo reports fewer error locations than \cpctplus, that
is probably due to \corchuelo's higher failure rate; however, as we
shall see in Section~\ref{sec:skipping input}, panic mode's much higher number of
error locations relative to \cpctplus might better be explained by other
factors.

With that caution in mind, the overall conclusions are fairly clear. \cpctplus
is able to repair nearly
all input files within the 0.5s timeout. While panic mode is able to repair
every file within the 0.5s timeout, it reports well over twice as many error
locations as \cpctplus -- in other words, panic mode substantially worsens the cascading
error problem. As well as producing more detailed and accurate output, \cpctplus has a
lower failure rate, median, and mean time than \corchuelo. The fact that the median
recovery time for \cpctplus is two orders of magnitude lower than its mean
recovery time suggests that only a small number of outliers cause error recovery to
take long enough to be perceptible to humans; this is confirmed by the
histogram in Figure~\ref{fig:results:cpctplus_histogram}. These results strongly
validate Hypothesis H1.

\corchuelo's poor performance may be surprising, as it produces at most
one (possibly non-minimum cost) repair sequence whereas \cpctplus produces the complete set of minimum
cost repair sequences -- in other words, \cpctplus is doing more work, more
accurately, and in less time than \corchuelo. There are three main reasons for \corchuelo's
poor performance. First,
the use of \textsc{CR Shift 1} causes the search to miss intermediate nodes
that would lead to success being detected earlier. Second, the heuristics used to stop the
search from going too far (e.g.~limiting a repair sequence's number of inserts and
deletes) are not well-suited to a large grammar such as Java's:
the main part of the search often exceeds the timeout, leaving no
time for the fallback mechanism of panic mode to be used. Finally, \corchuelo
lacks configuration merging, causing it to perform needless duplicate work.

\begin{figure}[t]
\vspace{-10pt}
\includegraphics[scale=.7]{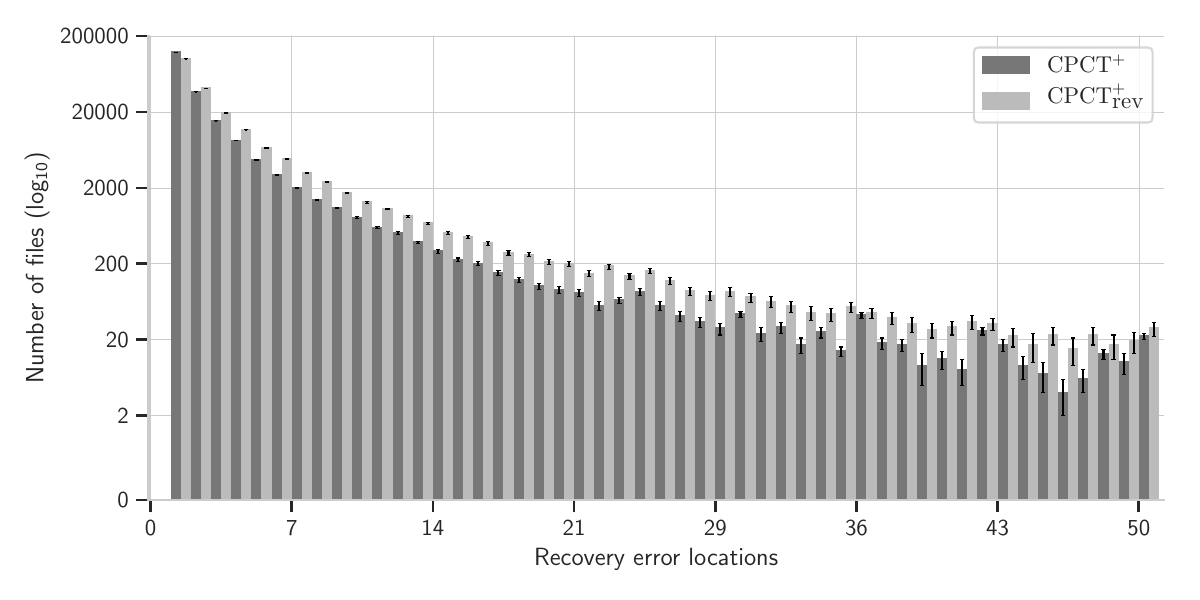}
\vspace{-25pt}
\caption{A histogram of the number of files with 1--50 error locations for
\cpctplus and \cpctplusrev. Note that we exclude: any files for which an error
recovery algorithms did not find a single error location (either because the
timeout was exceeded or no repair sequences could be found); and a
handful of outliers which distort the full histogram found in
Figure~\ref{fig:results:cpctplus_cpctplusrev_histogram_full} in the Appendix.
The $x$ axis shows the number of error
locations in a file and the $y$ axis is a logarithmic scale for the number of
files. Error bars represent 99\% confidence intervals. As this histogram
shows, the entire distribution is skewed
slightly rightwards by \cpctplusrev, showing that \cpctplusrev makes error recovery
slightly worse in a number of files (rather than making error recovery in a few
files a lot worse).}
\label{fig:results:cpctplus_cpctplusrev_histogram}
\end{figure}

\cpctplus ranks the complete set of minimum cost repair sequences by how
far parsing can continue and chooses from those which allow parsing to
continue furthest. \cpctplusrev, in contrast, selects from those which allow parsing
to continue the least far. \cpctplusrev shows that the ranking technique used in \cpctplus substantially
reduces the potential for cascading errors: \cpctplusrev leads to
\cpctplusreverrorlocsratioovercpctplus more error locations being reported to users
relative to \cpctplus. We visualise this in the histogram of
Figure~\ref{fig:results:cpctplus_cpctplusrev_histogram}
which shows all files with 1--50 error locations (a complete histogram can be
found in Figure~\ref{fig:results:cpctplus_cpctplusrev_histogram_full} in the
Appendix). Note that files where
error recovery did not complete and no error locations were found (which happens
occasionally with \cpctplusrev) are excluded from this histogram (since we know
that every file in the corpus has at least one error), but files where
error recovery did not complete but some error locations were found are
included (since this gives us, at least, a lower bound on the number of
error locations). As Figure~\ref{fig:results:cpctplus_cpctplusrev_histogram}
shows, the distribution of error locations in \cpctplus and \cpctplusrev is
similar, with the latter simply shifted slightly to the right.
In other words, \cpctplusrev
makes error recovery slightly worse in a number of files
(rather than making error
recovery in a small number of files a lot worse). This strongly validates
Hypothesis H2.

Interestingly, and despite its higher failure rate, \cpctplusrev has a
noticeably higher mean cost of repair sequences relative to
\cpctplus. In other words, \cpctplusrev not only causes more error locations to be reported,
but those additional error locations have longer repair sequences.  This
suggests that there is a double whammy from
cascading errors: not only are more error locations reported, but the poorer
quality repair sequences chosen make subsequent error locations
disproportionately harder for the error recovery algorithm to recover from.

\cpctplusdontmerge shows that configuration merging has a significant effect on
the failure rate, in our opinion justifying both its conceptual complexity and
the less than 100LoC Rust code taken to implement it. The slowdown
in the mean and median time for \cpctplusdontmerge suggests that configuration
merging is particularly effective on files with complex or numerous errors.

\subsection{The impact of skipping input}
\label{sec:skipping input}

The number of error locations reported
by panic mode is well over twice that of \cpctplus; even given \cpctplus's
higher failure rate relative to panic mode, this seemed hard to explain.
We thus made an additional hypothesis:

\begin{description}
  \item[H3] The greater the proportion of tokens that are skipped, the greater
    the number of error locations.
\end{description}

The intuition underlying this hypothesis is that, in general, the user's input
is very close to being correct and that the more input that error recovery skips,
the less likely it is to get back to a successful parse. We added the ability
to record the proportion of tokens skipped as the result of \emph{delete}
repairs during error recovery. The results in Figure~\ref{fig:results:summary}
show a general correlation between the
proportion of tokens skipped and the number of error locations (e.g.~\cpctplus
skips very little of the user's input; \cpctplusrev skips a little more; and
panic mode skips an order of magnitude more).
However, \corchuelo does not obviously follow this pattern: relative to the
other algorithms, its number of error locations does not correlate with the
proportion of input skipped. This is mostly explained by its
high mean time and high failure rate: \corchuelo tends to fail on files with
large numbers of error locations, underreporting the `true' number of error
locations simply because it cannot make it all of the way through such files
before the timeout. However, this outlier means that we consider
Hypothesis H3 to be only weakly validated.

\section{Using error recovery in practice}
\label{sec:api}

Although several members of the \fischer family were implemented in parsing
tools of the day, to the best of our knowledge none of those implementations
have survived. Equally, we are not aware of any member of the \fischer
family which explains how error recovery should be used or, indeed, if it has any
implications for users at all.

We are therefore forced to treat the following
as an open question: can one
sensibly use error recovery in the \fischer family in practice? In
particular, given that the most common way to use LR grammars
is to execute semantic actions as each production is reduced, what should semantic actions do
when parts of the input have been altered by error recovery? This latter
question is important for real-world systems (e.g.~compilers) which can
still perform useful computations (e.g.~running a type checker) in the
face of syntax errors.

While different languages are likely to lend themselves to different solutions, in
this section we show that \emph{grmtools} allows sensible integration
of error recovery in a Rust context. Readers who prefer to avoid
Rust-specific details may wish to move immediately to Section~\ref{sec:threats}.

\subsection{A basic solution}

\begin{figure}[!t]
\begin{adjustbox}{valign=t,minipage=.52\textwidth}
\begin{lstlisting}[numbers=left]
%start Expr
%%
Expr -> u64:
    Factor "+" Expr { $1 + $3 }
  | Factor { $1 }
  ;

Factor -> u64:
    Term "*" Factor { $1 * $3 }
  | Term { $1 }
  ;

Term -> u64:
    "(" Expr ")" { $2 }
  | "INT"
    {
      let n = $lexer.span_str($1.unwrap()
                                .span());
      match s.parse::<u64>() {
        Ok(val) => val as u64,
        Err(_) => panic!(
          "{} cannot be represented as a u64",
          s)
      }
    }
  ;
\end{lstlisting}
\end{adjustbox}
\hspace{3pt}
\begin{adjustbox}{valign=t,minipage=.46\textwidth}
\vspace*{0pt}
\caption{A naive version of the calculator grammar with semantic actions on each
production. The traditional Yacc \texttt{\%union} declaration is unwieldy in
Rust. Instead, \emph{grmtools} allows each rule to have a Rust return type
associated with it (between `\texttt{->}' and `\texttt{:}'): the actions of each
production in that rule must return values of that type. \texttt{\$\emph{n}}
variables reference the \emph{nth} symbol in a production (where
\texttt{\$1} references the first symbol). If that symbol is a reference to
a rule, a value of that rule's Rust type will be stored in that variable. If
that symbol is a token then the user's input can be obtained by
\texttt{\$lexer.span\_str(\$1.unwrap().span())} (line 17--18). Note that while this
grammar's semantic actions work as expected for inputs such as `2 + 3 * 4', they will
panic if too large a number is passed (lines 21--24), or if an integer is
inserted by error recovery. Figure~\ref{fig:api:resulttype} shows how to
avoid both problems.}
\label{fig:api:naive}
\end{adjustbox}
\end{figure}

Figure~\ref{fig:api:naive} shows a naive \emph{grmtools} version of the grammar
from Figure~\ref{fig:exprgrammar} that can evaluate numeric results as
parsing occurs (i.e.~given the input $2 + 3 * 4$ it returns $14$). This
grammar should mostly be familiar to Yacc users: each production has a \emph{semantic
action} (i.e.~Rust code that is executed when the production is reduced); and
symbols in the production are available to the semantic action
as pseudo-variables named \texttt{\$\emph{n}} (a production of
$n$ symbols has $n$ pseudo-variables with the first symbol connected to
\texttt{\$1} and so on). A minor difference from traditional Yacc is that
\emph{grmtools} allows rules to specify a different return type, an approach
shared with other modern parsers such as ANTLR~\cite{parr13definitive}.

A more significant difference relates to the \texttt{\$\emph{n}}
pseudo-variables: if they reference a rule $R$, then their type is
$R$'s return type; if they reference a token $T$, then their type
is (slightly simplified) \texttt{Result<Lexeme,
Lexeme>}. We will explain the reasons for this shortly, but at this stage it
suffices to note that, unless a token was inserted by error recovery,
we can extract tokens by calling \texttt{\$1.unwrap()}, and obtain the actual string the user passed by
using the globally available \texttt{\$lexer.span\_str} function.

\subsection{Can semantic action execution continue in the face of error recovery?}

\begin{figure}[!t]
\begin{adjustbox}{valign=t,minipage=.54\textwidth}
\begin{lstlisting}[numbers=left]
%start Expr
%avoid_insert "INT"
%%
Expr -> Result<u64, Box<dyn Error>>:
   Factor "+" Expr { Ok($1? + $3?) }
 | Factor { $1 }
 ;

Factor -> Result<u64, Box<dyn Error>>:
   Term "*" Factor { Ok($1? * $3?) }
 | Term { $1 }
 ;

Term -> Result<u64, Box<dyn Error>>:
  '(' Expr ')' { $2 }
 | 'INT'
   {
     let t = $1.map_err(|_|
       "<evaluation aborted>")?;
     let n = $lexer.span_str(t.span());
     match s.parse::<u64>() {
       Ok(val) => Ok(val as u64),
       Err(_) => {
         Err(Box::from(format!(
           "{} cannot be represented as a u64",
           s
         )))
       }
     }
   }
 ;
\end{lstlisting}
\end{adjustbox}
\hspace{3pt}
\begin{adjustbox}{valign=t,minipage=.44\textwidth}
\vspace*{0pt}
\caption{A more sophisticated version of the grammar from
Figure~\ref{fig:api:naive}. Each rule now returns a \texttt{Result} type. If an
integer is inserted by error recovery, the \texttt{Term} rule stops
evaluation by percolating the \texttt{Err} value upwards using the
`\texttt{?}' operator (which, if the \texttt{Result}-returning expression it is
attached to evaluates to an \texttt{Err}, immediately returns that error;
otherwise it unwraps the \texttt{Ok}); all other rules percolate such errors
upwards similiarly. As a convenience for the user, the contents of the `Err'
value are changed from a lexeme to a string explaining why the calculator
has not produced a value (lines 18-19). Note that other token types
are unaffected: if error recovery inserts a bracket, for example, evaluation of
the expression continues.}
\label{fig:api:resulttype}
\end{adjustbox}
\end{figure}

In Yacc, semantic actions can assume that each symbol in the
production has `normal' data attached to it (either a rule's value or the
string matching a token; Yacc's error recovery is implicitly expected to
maintain this guarantee) whereas, in our setting, inserted tokens
have a type but no value. Given the input `(2 + 3', the inserted close bracket is not
hugely important, and our calculator returns the value 5. However,
given the input `2 +',
\cpctplus finds a single repair sequence \emph{[Insert Int]}: what should
a calculator do with an inserted integer? Our naive calculator simply \texttt{panic}s
(which is roughly equivalent to `raises an exception and then exits') in such a
situation (the \texttt{unwrap} in Figure \ref{fig:api:naive} on line 17).
However, there are two alternatives to this rather extreme outcome:
the semantic action can assume a default value or stop further execution
of semantic values while allowing parsing to continue. Determining which is the
right action in the face of inserted tokens is inherently situation specific.
We therefore need a pragmatic way for users to control what happens in
such cases.

The approach we take is to allow users to easily differentiate normal vs.~inserted
tokens in a semantic action. Pseudo-variables that reference tokens have
(slightly simplified) the Rust type \texttt{Result<Lexeme, Lexeme>}. Rust's \texttt{Result}
type\footnote{Equivalents are found in several other languages: Haskell's
\texttt{Either}; O'Caml's \texttt{result}; or Scala's \texttt{Either}.} is a sum
type which represents success (\texttt{Ok($\ldots$)}) or error
(\texttt{Err($\ldots$)}) conditions. We use the \texttt{Ok} case
to represent `normal' tokens created from user input and the \texttt{Err} case
to represent tokens inserted by error recovery. Since the \texttt{Result} type
is widely used in Rust code, users can avail themselves of standard idioms.

For example, we can then alter our calculator grammar to continue
parsing, but stop executing meaningful semantic action code, when an inserted integer
is encountered. We change grammar rules from returning
type \texttt{T} to \texttt{Result<T, Box<dyn Error>{}>} (where \texttt{Box<dyn Error>}
is roughly equivalent to `can return any type of error'). It is then,
deliberately, fairly easy to use with the
\texttt{Result<Lexeme, Lexeme>} type: for tokens whose value we absolutely
require, we use Rust's `\texttt{?}' operator (which passes \texttt{Ok} values
through unchanged but returns \texttt{Err} values to the function caller) to
percolate our unwillingness to continue evaluation upwards. While \texttt{Box<dyn Error>} is
slightly verbose, it is a widely understood Rust idiom.
Figure~\ref{fig:api:resulttype} shows that changing the grammar to make use of
this idiom requires relatively little extra code.

\subsection{Avoiding insert repairs when possible}

Although we now have a reasonable mechanism for dealing with inserted tokens,
there are cases where we can bypass them entirely. For
example, consider the input `2 + + 3', which has two repair sequences
\emph{[Delete +], [Insert Int]}: evaluation of the expression can continue
with the former repair sequence, but not the latter. However, as
presented thus far, these repair sequences are ranked equally and one
non-deterministically selected.

We therefore added an optional declaration
\texttt{\%avoid\_insert} to \emph{grmtools} which allows users to specify
those tokens which, if inserted by error recovery, are likely to prevent
semantic actions from continuing execution. In practise, this is synonymous with those
tokens whose values (and not just their types) are important. In the calculator
grammar only the \texttt{INT} token satisfies this
criteria, so we add \texttt{\%avoid\_insert "INT"} to the grammar. We then
make a simple change to the repair sequence ranking of
Section~\ref{rankingrepairs} such that the final list of repair sequences is
sorted with inserts of such tokens at the bottom of the list. In our case, this
means that we always select \emph{Delete +} as the repair
sequence to apply to the input `2 + + 3' (i.e.~the
\emph{Insert Int} repair sequence is always presented as the second
option).

\subsection{Turning lexing errors into parsing errors}
\label{sec:lexing errors into parsing errors}

\begin{figure}[t]
\begin{minipage}{0.27\textwidth}
\begin{tabular}{p{0.05\textwidth}p{0.48\textwidth}}
\begin{subfigure}{0.02\textwidth}
\caption{}
\label{lst:lexerror:lexfile}
\end{subfigure}
&
\begin{minipage}[t]{0.83\textwidth}
\vspace{-13.5pt}
\begin{lstlisting}[language=Java]
. "ERRORTOKEN"
\end{lstlisting}
\end{minipage}
\\
\begin{subfigure}{0.02\textwidth}
\caption{}
\label{lst:lexerror:yaccfile}
\end{subfigure}
&
\begin{minipage}[t]{0.83\textwidth}
\vspace{-13.5pt}
\begin{lstlisting}
ErrorRule -> ():
  "ERRORTOKEN" { } 
  ;
\end{lstlisting}
\end{minipage}
\end{tabular}
\end{minipage}
\begin{minipage}{0.70\textwidth}
\vspace{-2.5pt}
\begin{tabular}{p{0.02\textwidth}p{0.48\textwidth}}
\begin{subfigure}{0.02\textwidth}
\caption{}
\label{lst:lexerror:output}
\end{subfigure}
&
\begin{minipage}[t]{0.93\textwidth}
\vspace{-13.5pt}
\begin{lstlisting}
Parsing error at line 1 column 3. Repair sequences found:
   1: Insert +, Delete @
   2: Insert *, Delete @
   3: Delete @, Delete 3
Result: 9
\end{lstlisting}
\end{minipage}
\end{tabular}
\end{minipage}
\vspace{-10pt}
\caption{A simple way of turning lexing errors into parsing errors in
\emph{grmtools}. First, we add a \texttt{ERRORTOKEN} token type, which matches otherwise invalid input,
to the end of the Lex file (\subref{lst:lexerror:lexfile}). Second, we add a
rule \texttt{ErrorRule} to the Yacc grammar referencing \texttt{ERRORTOKEN}
(\subref{lst:lexerror:yaccfile}). Note that \texttt{ErrorRule} must not be
referenced by any other rule in the Yacc grammar. With those two steps
complete, input with lexing errors such as `\texttt{2 @ 3 + 4}' invokes normal error
recovery (\subref{lst:lexerror:output}).}
\label{lst:lexerror}
\end{figure}

In most traditional parsing systems, lexing errors are distinct from parsing
errors: only files which can be fully lexed can be parsed. This is confusing
for users, who are often unaware of the distinction between these two phases.
To avoid this, we lightly adapt the idea of \emph{error tokens} from
incremental parsing~\cite[p.~99]{wagner98practicalalgorithms}. In essence, any
input which cannot be lexed is put into a token whose type is not referenced in
the normal grammar. This guarantees that all possible input lexes without
error and, when the parser encounters an error token, normal error recovery
commences\footnote{Note that this is an opt-in feature: it was not enabled
for the experiments in Section~\ref{experiment}.}.

A basic, but effective, version of this requires no special support from
\emph{grmtools} (see Figure~\ref{lst:lexerror}). First, we add a new toke type to
the lexer that matches each otherwise invalid input character.
Second, since \emph{grmtools} requires that all tokens defined in the lexer are
referenced in the grammar, we add a dummy rule to the grammar that references
the token (making sure not to reference this rule elsewhere in the grammar).
These two steps are sufficient to ensure that users always see the same style
of error messages, and the same style of error recovery, no matter whether they
make a lexing or a parsing error.

\section{Threats to validity}
\label{sec:threats}

Although it might not be obvious at first, \cpctplus is
non-deterministic, which can lead to
different results from one run to the next. The root cause of this problem is
that multiple repair sequences may have identical effects up to
$N_\textit{try}$ tokens, but cause different effects after that value.
By running each file through each
error recovery multiple times and reporting confidence intervals, we are able
to give a good -- though inevitably imperfect -- sense of the likely variance
induced by this non-determinism.

Our implementation of \corchuelo is a `best effort'. The description in the
paper is incomplete in places and, to the best of our knowledge, the
accompanying source code is no longer available. We thus may not have
faithfully implemented the intended algorithm.

Blackbox contains an astonishingly large amount of source code but has two
inherent limitations. First, it only contains Java source code. This means that
our main experiment is limited to one grammar: it is possible that our
techniques do not generalise beyond the Java grammar (though, as
Appendix~\ref{app:examples} suggests, our techniques do appear to work well
on other grammars). Although \cite[p.~109]{cerecke03phd}
suggests that different grammars make relatively little difference
to the performance of such error recovery algorithms, we are not aware
of an equivalent repository for other language's source code. One solution is
to mutate correct source files (e.g.~randomly deleting tokens), thus
obtaining incorrect inputs which we can later test: however, it is difficult
to uncover and then emulate the numerous, sometimes surprising, ways that
humans make syntax errors, particularly as some are language specific
(though there is some early work in this area~\cite{dejonge12automated}).
Second, Blackbox's data comes largely from students,
who are more likely than average to be somewhat novice programmers. It is clear
that novice programmers make some different syntax errors -- and, probably, make
some syntax errors more often -- relative to advanced programmers. For example,
many of the files with the greatest number of syntax errors are caused by
erroneous fragments repeated with variants (i.e.~it is likely that the
programmer wrote a line of code, copy and pasted it, edited it, and repeated
that multiple times before deciding to test for syntactic validity). It is
thus possible that a corpus consisting solely of programs from advanced programmers
would lead to different results. We consider this a minor worry,
partly because a good error recovery algorithm should
aim to perform well with inputs from users of different experience levels.

Our corpus was parsed using a Java 1.7 grammar, but some members of the corpus
were almost certainly written using Java 1.8 or later features. Many -- though not all -- post-1.7 Java
features require a new keyword: such candidate source files would thus have
failed our initial lexing test and not been included in our corpus. However,
some Java 1.8 files will have made it through our checks. Arguably these are still a valid
test of our error recovery algorithms. It is even likely that they may be a little
more challenging on average, since they are likely to be further away from being valid
syntax than files intended for Java 1.7.

\section{Related work}

Error recovery techniques are so numerous that there is no
definitive reference or overview of them. However, \cite{degano95comparison}
contains an overall historical analysis and \cite{cerecke03phd} an excellent
overview of much of the \fischer family. Both
must be supplemented with more recent works.

The biggest limitation of error recovery algorithms in the
\fischer family (including \cpctplus) is that they find repairs at the
point that an error is discovered, which may be later in the file than the cause
of the error. Thus even when they successfully recover from an error, the repair
sequence reported may be very different from the fix the user considers
appropriate (note that this is distinct from the cascading error problem,
which our ranking of repair sequences in Section~\ref{rankingrepairs} partly
addresses). A common, frustrating, example of this is a missing `\texttt{\}}' character in
C/Java-like languages. Some approaches are able to backtrack from the source of
the error in order to try and find more appropriate repairs. However, there are
two challenges to this: first, the cost of maintaining the necessary state to
backtrack slows down normal parsing (e.g.~\cite{deJonge12natural} only stores the
relevant state at each line encountered to reduce this cost), whereas we add no
overhead at all to normal parsing; second, the search-space is so hugely
increased that it can be harder to find any repairs at
all~\cite{degano95comparison}.

One approach to global error recovery is to use machine
learning to train a system on syntactically correct programs~\cite{santos18syntax}: when a
syntax error is encountered, the resulting model is used to suggest appropriate global
fixes. Although \cite{santos18syntax} also use data from Blackbox, their
experimental methodology is both stricter -- aiming to find exactly the same
repair as the human user applied -- and looser -- they only
consider errors which can be fixed by a single token, discarding 42\% of
the data~\cite[p.~8]{santos18syntax}) whereas we attempt to fix errors which
span multiple tokens. It is thus difficult to directly compare our results to
theirs. However, by the high bar they have set themselves, they are able to repair
52\% of single-token errors.

As \cpctplusrev emphasises, choosing an inappropriate repair sequence during
error recovery leads to cascading errors. The noncorrecting error recovery approach
proposed by~\cite{richter85noncorrecting} explicitly addresses this weakness,
eschewing repairs entirely. When a syntax error is discovered, noncorrecting
error recovery attempts to discover all further syntax errors by checking
whether the remaining input (after the point an error is detected) is a valid suffix in the
language. This is achieved by creating a recogniser that can identify all valid
suffixes in the language. Any errors identified in the suffix parse are
guaranteed to be genuine syntax errors because they are uninfluenced by
errors in the (discarded) prefix (though this does mean that some
genuine syntax errors are missed that would not have been valid
suffixes at that point in the user's input had the original syntax error not
been present). There seem to be two
main reasons why noncorrecting error recovery has
not been adopted. First, building an appropriate recogniser is surprisingly
tricky and we are not currently aware of one that can handle the full class of
LR grammars (though the full class of LL grammars has been
tackled~\cite{deudekom93initial}), though we doubt that this problem is
insoluble. Second, as soon as a syntax error is encountered, noncorrecting error
recovery is unable to execute semantic actions, since it lacks the execution context
they need.

Although one of our paper's aims is to find the complete set of minimum cost repair sequences,
it is unclear how best to present them to users, leading to questions such as:
should they be simplified? should a subset be presented? and so on. Although
rare, there are some surprising edge cases. For example,
the (incorrect) Java 1.7 expression `\texttt{x = f(""a""b);}' leads to 23,067 minimum
cost repair sequences being found, due to the large number of Java keywords that are
valid in several parts of this expression leading to a combinatorial explosion: even the most
diligent user is unlikely to find such a volume of information valuable. In
a different vein, the success condition of `reached an accept' state is
encountered rarely enough that we sometimes forgot that it could happen and
were confused by an apparently unexplained difference in the repair sequences
reported for the same syntax chunk when it was moved from the end to the middle
of a file. There is a
body of work which has tried to understand how best to structure compiler error
messages (normally in the context of those learning to program). However, the
results are hard to interpret: some studies find that more complex error
messages are not useful~\cite{nienaltowski08error}, while others suggest they
are~\cite{prather17novices}. It is unclear to us what the right approach might be,
or how it could be applied in our context.

The approach of \cite{mckenzie95error} is similar to
\corchuelo, although the former cannot incorporate shift
repairs. It tries harder than \cpctplus to prune out pointless search
configurations~\cite[p.~12]{mckenzie95error}, such as cycles in the parsing stack,
although this leads to some minimum cost repairs being
skipped~\cite{bertsch99failure}. A number of interlocking, sophisticated pruning
mechanisms which build on this are described in~\cite{cerecke03phd}. These are
significantly more complex than our merging of compatible configurations: since this
gives us acceptable performance in practise, we have not investigated other
pruning mechanisms.

The most radical member of the \fischer family is that
of~\cite{kimyi10astar}\footnote{In an earlier online draft of this paper we
stated that this algorithm has a fundamental flaw. We now believe this was
due to us incorrectly assuming that the `delete' optimisation of \corchuelo
applied to \cite{kimyi10astar}. We apologise to the authors for this
mistake.}. This generates repair sequences in the vein of \corchuelo
using the A* algorithm and a precomputed distance table.
\cite{kimyi10astar} works exclusively on the stategraph, assuming that it is
unambiguous. However, Yacc systems allow ambiguous stategraphs and provide a means for
resolving those ambiguities when creating the statetable. Many real-world
grammars (e.g.~Lua, PHP) make use of ambiguity resolution. In
an earlier online draft, we created \mf, a statetable-based algorithm which
extends \cpctplus with ideas from \cite{kimyi10astar} at the cost of
significant additional complexity. With the benefit of hindsight, we do not
consider \mf's relatively small benefits (e.g.~reducing the failure rate by
approximately an additional 0.5\%) to be worth that
additional complexity.

\cpctplus takes only the grammar and token types into account. However,
it is possible to use additional information, such as nesting (e.g.~taking into
account curly brackets) and indentation when recovering from errors. This
has two aims: reducing the size of the search space (i.e.~speeding up error
recovery); and making it more likely that the repairs reported matched
the user's intentions. The most sophisticated
approach in this vein we are aware of is that of~\cite{deJonge12natural}. At
its core, this approach uses GLR parsing: after a grammar is suitably
annotated by the user, it is then transformed into a `permissive' grammar which
can parse likely erroneous inputs; strings which match the permissive parts of
the grammar can then be transformed into a non-permissive counterpart. In
all practical cases, the transformed grammar will be ambiguous, hence the
need for generalised parsing. Our use of parent-pointer trees
in configuration merging gives that part of our algorithm a similar feel to GLR parsing (even though
we do not generate ambiguous strings). However, there are major
differences: LR parsers are much simpler than GLR parsers;
and the \fischer family of algorithms do not require manually
annotating, or increasing the size of, the grammar.

A different approach to error recovery is that taken by~\cite{pottier16reachability}: rather
than try and recover from errors directly, it reports in natural language
how the user's input caused the parser to reach an error state (e.g.~``I
read an open bracket followed by an expression, so I was expecting a close
bracket here''), and
possible routes out of the error (e.g.~``A function or variable declaration is
valid here''). This involves significant manual work, as every parser state
(1148 in the Java grammar we use) in which an error can occur needs to be
manually marked up, though the approach has
various techniques to lessen the problem of maintaining messages as a grammar
evolves.

Many compilers and editors have hand-written parsers with hand-written error
recovery. Though generally ad-hoc in their approach, it is possible, with
sufficient effort, to make them perform well. However, this comes at
a cost. For example, the hand-written error recovery routines in the Eclipse IDE
are approximately 5KLoC and are solely for use with Java code: \cpctplus
is approximately 500LoC and can be applied to any LR grammar.

Although error recovery approaches have, historically, been mostly LR based,
there are several non-LR approaches. A full overview is impractical, though a
few pointers are useful. When LL parsers encounter an error, they generally
skip input until a token in the follow set is encountered (an early example
is~\cite{turner77error}). Although this outperforms the simple panic mode of
Section~\ref{sec:panic mode}, it will, in general, clearly skip more input than
\cpctplus, which is undesirable. LL parsers do, however, make it
somewhat easier to express grammar-specific error recovery rules.
The most advanced LL approach that we are aware of is
IntelliJ's Grammar-Kit, which allows users to annotate their grammars for error
recovery. Perhaps the most interesting annotation is that certain rules can be
considered as fully matched even if only a prefix is matched (e.g.~a partially
completed function is parsed as if it was complete). It might be possible to add
similar ideas to a successor of \cpctplus, though this is more
awkward to express in an LR approach. Error recovery for PEG grammars is much more
challenging, because the non-LL parts of the grammar mean that there is not
always a clearly defined point at which an error is determined to have
occurred. PEG error recovery has thus traditionally required extensive manual
annotations in order to achieve good quality recovery. \cite{medeiros19syntax}
tackles this problem by
automatically adding many (though not necessarily all) of the annotations
needed for good PEG error recovery. However, deciding when to add, and when not to
add, annotations is a difficult task and the two algorithms presented have
different trade-offs: the \emph{Standard} algorithm adds more annotations,
leading to better quality error recovery, but can change
the input language accepted; the \emph{Unique}
algorithm adds fewer annotations, leading to poorer quality error recovery, but
does not affect the language accepted. The quality of error recovery of the
Unique algorithm, in particular, is heavily dependent on the input grammar: it works
well on some (e.g.~Pascal) but less well on others (e.g.~Java). In cases
where it performs less well, it can lead to parsers which skip large portions
(sometimes the remainder) of the input.

While the field of programming languages has largely forgotten the approach of
\cite{aho72minimum}, there are a number of successor works (e.g.~\cite{rajasekaran16error}).
These improve the time complexity, though none
that we are aware of address the issue of how to present to the user what has been done.

We are not aware of any error recovery algorithms that are formally verified.
Indeed, as shown in this paper, some have serious flaws. We are only aware of
two works which have begun to consider what correctness for such algorithms might
mean:~\cite{zaytsev14formal} provides a brief philosophical justification of the need
and~\cite{gomezrodriguez10error} provides an outline of an approach. Until
such time as someone verifies a full error recovery algorithm, it is difficult
to estimate the effort involved, or what issues may be uncovered.

\section{Conclusions}

In this paper we have shown that error recovery algorithms in the
\fischer family can run fast enough to be usable in the real
world. Extending such algorithms to produce the complete set of
minimum cost repair sequences allows parsers to provide better feedback to
users, as well as significantly reducing the cascading error problem.
The \cpctplus algorithm is simple to implement (less than 500LoC in our
Rust system) and still has good performance.

Looking to the future, we, perhaps immodestly, suggest that \cpctplus might be `good enough'
to serve as a common representative of the \fischer family.
However, we do not think that it is the perfect solution. We suspect
that, in the future, multi-phase solutions will be developed. For example, one
may use noncorrecting error recovery (e.g.~\cite{richter85noncorrecting})
to identify syntax errors, and then use a combination of machine-learning
(e.g.~\cite{santos18syntax}) and \cpctplus to discover those repair sequences
that do not lead to additional error locations being encountered.

\textbf{Acknowledgements:}
We are grateful to the Blackbox developers for allowing us access
to their data, and particularly to Neil Brown for help in extracting a relevant
corpus. We thank Edd Barrett for helping to set up our benchmarking
machine and for comments on the paper. We also thank Carl Friedrich
Bolz-Tereick, Sérgio Queiroz de Medeiros, Sarah Mount, François Pottier, Christian Urban,
and Naveneetha Vasudevan for comments.
This research was funded by the EPSRC Lecture (EP/L02344X/1) Fellowship.

\bibliographystyle{plain}
\bibliography{bib}

\appendix

\section*{Appendix (not peer-reviewed)}

\section{Curated examples}
\label{app:examples}

In this section we show several examples of error recovery using \cpctplus in
different languages, to give some idea of what error recovery looks like in
practise, and to emphasise that the algorithms in this paper are grammar
neutral. All of these examples use the output from the \emph{nimbleparse}
tool that is part of \emph{grmtools}.

\subsection{Java 7}

\noindent Example 1 input:
\begin{lstlisting}[numbers=left,language=Java,xleftmargin=2em]
class C {
  int x y;
}
\end{lstlisting}

\noindent Example 1 output:
\begin{lstlisting}[xleftmargin=2em]
Parsing error at line 2 column 9. Repair sequences found:
   1: Delete y
   2: Insert ,
   3: Insert =
\end{lstlisting}

\vspace{12pt}

\noindent Example 2 input:
\begin{lstlisting}[numbers=left,language=Java,xleftmargin=2em]
class C {
  void f() {
    if true {
  }
}
\end{lstlisting}

\noindent Example 2 output:
\begin{lstlisting}[xleftmargin=2em]
Parsing error at line 3 column 8. Repair sequences found:
   1: Insert (, Shift true, Insert )
Parsing error at line 5 column 2. Repair sequences found:
   1: Insert }
\end{lstlisting}

\vspace{12pt}

\noindent Example 3 (taken from \cite[p.~10]{deJonge12natural}) input:
\begin{lstlisting}[numbers=left,language=Java,xleftmargin=2em]
class C {
  void f() {
    if (temp.greaterThan(MAX) // missing )
      fridge.startCooling();
  }
}
\end{lstlisting}

\noindent Example 3 output:
\begin{lstlisting}[xleftmargin=2em]
Parsing error at line 4 column 7. Repair sequences found:
   1: Insert )
\end{lstlisting}

\vspace{12pt}

\noindent Example 4 (taken from \cite[p.~16]{deJonge12natural}) input:
\begin{lstlisting}[numbers=left,language=Java,xleftmargin=2em]
class C {
  void methodX() {
    if (true)
      foo();
    }
    int i = 0;
    while (i < 8)
      i=bar(i);
    }
  }
}
\end{lstlisting}

\noindent Example 4 output:
\begin{lstlisting}[xleftmargin=2em]
Parsing error at line 7 column 5. Repair sequences found:
   1: Insert {
Parsing error at line 11 column 1. Repair sequences found:
   1: Delete }
\end{lstlisting}

\vspace{12pt}

\noindent Example 5 (taken from \cite[p.~2]{medeiros18syntax}):
\begin{lstlisting}[numbers=left,language=Java,xleftmargin=2em]
public class Example {
  public static void main(String[] args) {
    int n = 5;
    int f = 1;
    while(0 < n) {
      f = f * n;
      n = n - 1
    };
    System.out.println(f);
  }
}
\end{lstlisting}

\noindent Example 5:
\begin{lstlisting}[xleftmargin=2em]
Parsing error at line 8 column 5. Repair sequences found:
   1: Delete }
   2: Insert ;
Parsing error at line 11 column 2. Repair sequences found:
   1: Insert }
\end{lstlisting}

\vspace{12pt}

\noindent Example 6:
\begin{lstlisting}[numbers=left,language=Java,xleftmargin=2em]
class C {
  void f() {
    x((((((
  }
}
\end{lstlisting}

\noindent Example 6 output, showing the timeout being exceeded and error recovery
unable to complete:
\begin{lstlisting}[xleftmargin=2em]
Parsing error at line 4 column 3. No repair sequences found.
\end{lstlisting}

\subsection{Lua 5.3}

\noindent Example 1 input:
\begin{lstlisting}[numbers=left,xleftmargin=2em]
print("Hello World"
\end{lstlisting}

\noindent Example 1 output:
\begin{lstlisting}[xleftmargin=2em]
Parsing error at line 1 column 20. Repair sequences found:
   1: Insert )
\end{lstlisting}

\vspace{12pt}

\noindent Example 2 input. Note that `\texttt{=}' in Lua is the assignment
operator, which is not valid in conditionals; and that if/then/else blocks must
be terminated by `\texttt{end}'.
\begin{lstlisting}[numbers=left,xleftmargin=2em]
function fact (n)
  if n = 0 then
    return 1
  else
    return n * fact(n-1)
end
\end{lstlisting}

\noindent Example 2 output:
\begin{lstlisting}[xleftmargin=2em]
Parsing error at line 2 column 8. Repair sequences found:
    1: Delete =, Delete 0
    2: Insert or, Delete =
    3: Insert and, Delete =
    4: Insert ==, Delete =
    5: Insert ~=, Delete =
    6: Insert >=, Delete =
    7: Insert <=, Delete =
    8: Insert >, Delete =
    9: Insert <, Delete =
   10: Insert |, Delete =
   11: Insert ~, Delete =
   12: Insert &, Delete =
   13: Insert >>, Delete =
   14: Insert <<, Delete =
   15: Insert -, Delete =
   16: Insert +, Delete =
   17: Insert .., Delete =
   18: Insert %, Delete =
   19: Insert //, Delete =
   20: Insert /, Delete =
   21: Insert *, Delete =
   22: Insert ^, Delete =
Parsing error at line 6 column 4. Repair sequences found:
   1: Insert end
\end{lstlisting}

\vspace{12pt}

\noindent Examples 3 and 4 (both derived from the Lua 5.3 reference manual) show
that \cpctplus naturally deals with an inherent ambiguity in Lua's Yacc
grammar involving function calls and assignments (which, following the Lua
specification, is resolved by Yacc in favour of function calls). Example 3
shows the `unambiguous' case (i.e.~if Lua forced users to use `;'
everywhere, the grammar would have no ambiguities):

\begin{lstlisting}[numbers=left,xleftmargin=2em]
a = b + c;
(print or io.write)'done')
\end{lstlisting}

\noindent Example 3 output:
\begin{lstlisting}[xleftmargin=2em]
Parsing error at line 2 column 26. Repair sequences found:
   1: Delete )
   2: Insert (
\end{lstlisting}

\vspace{12pt}

\noindent Example 4 shows what happens in the `ambiguous' case (which Lua's
grammar resolves in favour of viewing the code below as a function call to
\texttt{c}):

\begin{lstlisting}[numbers=left,xleftmargin=2em]
a = b + c
(print or io.write)'done')
\end{lstlisting}

\noindent Example 4 output:
\begin{lstlisting}[xleftmargin=2em]
Parsing error at line 2 column 26. Repair sequences found:
   1: Delete )
\end{lstlisting}

\vspace{12pt}

\noindent Example 5 (taken from \cite[p.~7]{medeiros18syntax}):

\begin{lstlisting}[numbers=left,xleftmargin=2em]
if then print("that") end
\end{lstlisting}

\noindent Example 5 output:
\begin{lstlisting}[xleftmargin=2em]
Parsing error at line 1 column 4. Repair sequences found:
   1: Insert [=[<String>]=]
   2: Insert "<String>"
   3: Insert ...
   4: Insert <Numeral>
   5: Insert true
   6: Insert false
   7: Insert nil
   8: Insert <Name>
\end{lstlisting}

\subsection{PHP 7.3}

\noindent Example 1 input:
\begin{lstlisting}[numbers=left,xleftmargin=2em]
function n() {
    $x = 1
}
\end{lstlisting}

\noindent Example 1 output:
\begin{lstlisting}[xleftmargin=2em]
Parsing error at line 3 column 1. Repair sequences found:
   1: Insert ;
\end{lstlisting}

\vspace{12pt}

\noindent Example 2 input:
\begin{lstlisting}[numbers=left,xleftmargin=2em]
$a = array("foo", "bar");
$a[0;
\end{lstlisting}

\noindent Example 2 output:
\begin{lstlisting}[xleftmargin=2em]
Parsing error at line 2 column 5. Repair sequences found:
   1: Insert ]
\end{lstlisting}

\vspace{12pt}

\noindent Example 3 input:
\begin{lstlisting}[numbers=left,xleftmargin=2em]
class X {
    function a($x) {
        if $x {
    }
}
\end{lstlisting}

\noindent Example 3 output:
\begin{lstlisting}[xleftmargin=2em]
Parsing error at line 3 column 12. Repair sequences found:
   1: Insert (, Shift $x, Insert )
Parsing error at line 5 column 2. Repair sequences found:
   1: Insert }
\end{lstlisting}

\begin{figure}[t]
\includegraphics[scale=.7]{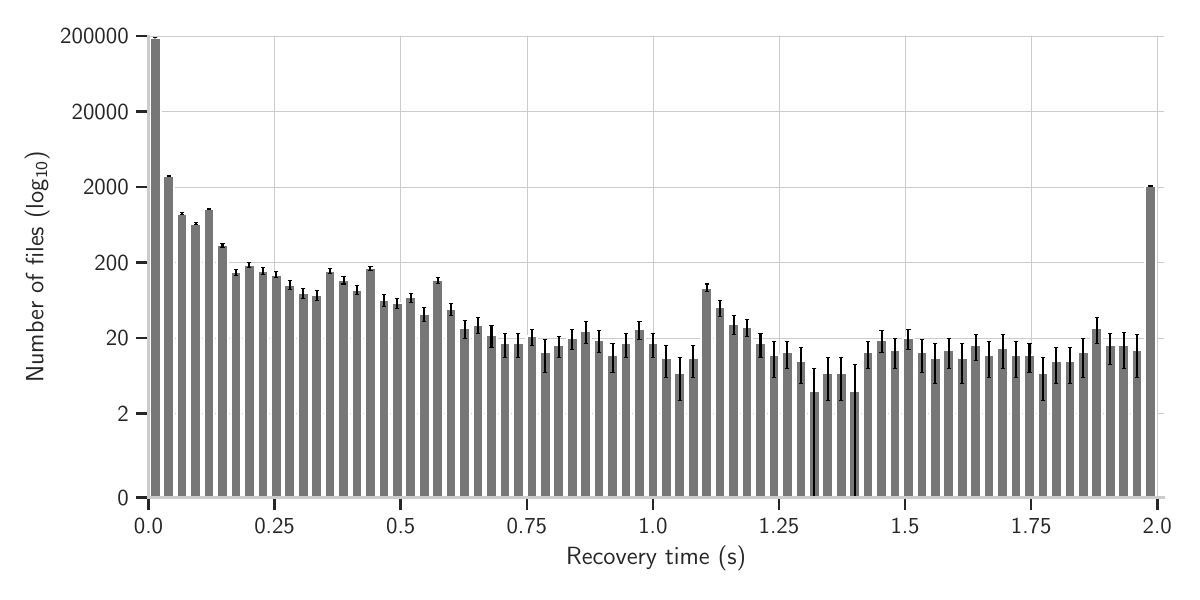}
\vspace{-25pt}
\caption{A histogram showing the effect of increasing the timeout from 0.5s to
2s (see Figure~\ref{fig:results:cpctplus_histogram} for the histogram showing a
timeout of 0.5s). Increasing the timeout from 0.5s to 2s lowers the failure
rate from \cpctplusfailurerate to \cpctpluslongerfailurerate, with a slowly
decreasing number of files succeeding as the timeout increases.
Although we used a timeout of 0.5s on the basis that we felt most users
would tolerate such a delay, others may wish to pick a shorter or longer
timeout depending on their perception of their users tolerance of delay
vs.~tolerance of failed error correction.}
\label{fig:results:cpctplus_longer_histogram}
\end{figure}

\begin{figure}[t]
\includegraphics[scale=.7]{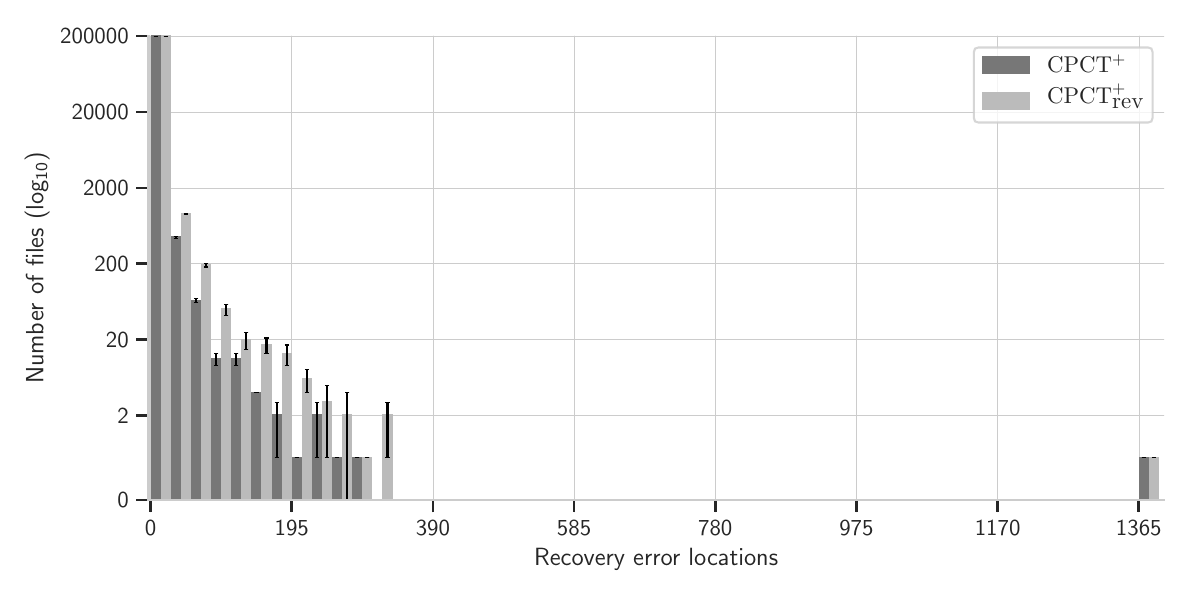}
\vspace{-25pt}
\caption{The full histogram of the number of error locations. The small number
of outliers obscures the main bulk of the data -- see
Figure~\ref{fig:results:cpctplus_cpctplusrev_histogram} for the truncated
version.}
\label{fig:results:cpctplus_cpctplusrev_histogram_full}
\end{figure}

\end{document}